\documentclass{emulateapj}
\usepackage{natbib,graphicx,subfigure}
\slugcomment{{\sc Accepted for Publication in the Astronomical Journal:} March 25, 2008}
\shorttitle{FIRST/SDSS BL Lacs}
\shortauthors{Plotkin et al.}

\begin{document}
\newcommand{\nsamp}{501}    %Total # of objects in sample
\newcommand{\nbl}{426}           %# of BL objects
\newcommand{\nmaybe}{75}    %# of BL? objects
%%%%%same definitions for objects with X-ray matches
\newcommand{\nsampxray}{230}
\newcommand{\nblxray}{217}
\newcommand{\nmaybexray}{13}
\tabletypesize{\scriptsize}

\title{A Large Sample of BL Lacs from SDSS and FIRST}

\author{
Richard~M.~Plotkin,\altaffilmark{1} 
Scott~F.~Anderson,\altaffilmark{1} 
Patrick~B.~Hall,\altaffilmark{2}
Bruce~Margon,\altaffilmark{3}
Wolfgang~Voges,\altaffilmark{4}
Donald~P.~Schneider,\altaffilmark{5}
Gregory~Stinson,\altaffilmark{1,6}
and Donald~G.~York\altaffilmark{7}
}

\altaffiltext{1}{Department of Astronomy, University of Washington, Box 351580, Seattle, WA 98195, USA; plotkin@astro.washington.edu, anderson@astro.washington.edu}
\altaffiltext{2}{Department of Physics and Astronomy, York University, 4700 Keele Street, Toronto, ON M3J 1P3, Canada.}
\altaffiltext{3}{Department of Astronomy and Astrophysics, University of California, 1156 High Street, Santa Cruz, CA 95064, USA}
\altaffiltext{4}{Max-Planck-Institut f\"{u}r extraterrestrische Physik, Giessenbachstrasse, Postfach 1312, D-85741, Garching, Germany}
\altaffiltext{5}{Department of Astronomy and Astrophysics, Pennsylvania Sate University, 525 Davey Laboratory, University Park, PA 16802, USA}
\altaffiltext{6}{Department of Physics and Astronomy, McMaster University, Hamilton, ON L8S 4M1, Canada}
\altaffiltext{7}{Department of Astronomy and Astrophysics and the Enrico Fermi Institute, 5640 South Ellis Avenue, Chicago, IL 60637, USA}

%%%%%%%%%%%%%%%%%%%%%%%%%%%%%%%
\begin{abstract}
We present a large sample of \nsamp\ radio-selected BL Lac candidates from the combination of the Sloan Digital Sky Survey (SDSS) Data Release 5 optical spectroscopy and from the Faint Images of the Radio Sky at Twenty-Centimeters (FIRST) radio survey; this is one of the largest BL Lac samples yet assembled, and each object emerges with homogeneous data coverage.  Each candidate is detected in the radio from FIRST and confirmed in SDSS optical spectroscopy to have: (1) no emission feature with measured rest equivalent width larger than 5~\AA; and (2) no measured \ion{Ca}{2} H/K depression larger than 40\%.  We subdivide our sample into \nbl\ higher confidence candidates and \nmaybe\ lower confidence candidates.  We argue that contamination from other classes of objects that formally pass our selection criteria is small, and we identify a few very rare radio AGN with unusual spectra that are probably related to broad absorption line quasars.  About one-fifth of our sample were known BL Lacs prior to the SDSS.  A preliminary analysis of the sample generally supports the standard beaming paradigm.  While we recover sizable numbers of low-energy and intermediate-energy cutoff BL Lacs (LBLs and IBLs, respectively), there are indications of a potential bias toward recovering high-energy cutoff BL Lacs (HBLs) from SDSS spectroscopy.  Such a large sample may eventually provide new constraints on BL Lac unification models and their potentially peculiar cosmic evolution;  in particular, our sample contains a significant number of higher redshift objects, a sub-population for which the standard paradigm has yet to be rigorously constrained.

\end{abstract}


%%%%%%%%%%%%%%%%%%%%%%%%
%%%%%%%%%%%%%%%%%%%%%%%%
\section{Introduction}
\label{sec:intro}
BL Lacs are a rare subclass of Active Galactic Nuclei (AGN) characterized by nearly featureless optical spectra, multiwavelength emission, marked variability and strong polarization \citep[e.g., see][]{blandford78,kollgaard94, urry95, perlman01}.  Their spectral energy distributions (SEDs) tend to be dominated by synchrotron radiation, with a component due to inverse Compton scattering at higher frequencies.  The standard paradigm describes BL Lacs in the context of AGN unification as FR I radio galaxies with their relativistic jets pointed toward the observer \citep[e.g.,][]{blandford78}.   In this scenario the observed jets are brightened through beaming effects, making BL Lacs powerful probes of  AGN jet physics and radio-loud AGN in general.  Their rarity and lack of strong spectral features historically render many traditional quasar selection techniques (e.g., UV-excess or objective prism surveys) inefficient.  Still, large BL Lac samples are eagerly sought but until relatively recently have remained elusive;  given their strong multiwavelength continuous emission and rarity, broad-band recovery methods over large regions of the sky tend to be the most efficient.

BL Lac studies are sensitive to selection biases and historically have been limited by small number statistics.   For example, a now disfavored dichotomy between radio selected BL Lacs (RBLs) and X-ray selected BL Lacs (XBLs) was initially suggested from a comparison between the small and relatively shallow but complete 1 Jy radio sample \citep{stickel91} and an X-ray selected sample from the {\it Einstein} Observatory Extended Medium-Sensitivity Survey \citep[EMSS;][]{stocke91}.  In addition to XBLs appearing to have less rapid variability and smaller nuclear to host galaxy flux ratios, RBLs and XBLs were thought to populate distinctly different regions of parameter space \citep[e.g., see Figure 5 of][]{padovani95_apj}.  	XBLs are also more weakly polarized on average, and the majority of XBLs have constant polarization position angles (within 20$^{\circ}$) over timescales of at least 3 years \citep{jannuzi94}. 

But \citet{padovani95_apj}~argue against two distinct BL Lac populations, and they instead propose that the observed differences between RBLs and XBLs could be explained via different energy cutoffs in the beamed synchrotron radiation component of their SEDs.  They suggest a more physical naming scheme of {\it low-energy cutoff BL Lac} (LBL) and {\it high-energy cutoff BL Lac} (HBL) to replace RBL and XBL.  LBLs have energy cutoffs in the near-infrared/optical, while HBLs have energy cutoffs in the UV/X-ray.  Most but not all RBLs are LBLs in this scenario, and similarly most XBLs are HBLs.    Deeper surveys have subsequently found many objects, aptly named {\it intermediate-energy cutoff BL Lacs} (IBL), that bridge this LBL/HBL gap \citep[e.g.,][]{kock96,perlman96,laurent99,collinge05}, thereby demonstrating the apparent dichotomy's  emergence as a selection effect.  However, any model advocating a single BL Lac population is still challenged with explaining the phenomenological differences between RBLs and XBLs described in the previous paragraph. 

Large and deep samples are clearly essential for advancing our knowledge on BL Lacs \citep[e.g., see][]{padovani95_apj,perlman01,chen_l.e.06}, and much progress has been made over the past 10-15 years: the number of known BL Lacs has ballooned from only a couple hundred \citep{padovani95_mnras} to about 1100 catalogued in \citet{veron06}.  Numerous selection techniques have been used to assemble sizable BL Lac samples.  One multifrequency selection approach identifies candidates based on flat radio spectra, followed by requisite optical spectroscopy \citep[e.g.,][]{kuehr90,stickel91,marcha96,caccianiga02_mnras}, while another popular technique assembles catalogs via the indentification of optical counterparts to radio and/or X-ray error circles in various surveys \citep[e.g.,][]{stocke91,wolter97,bade98,fischer98,laurent98,maccacaro98,laurent99,perlman98,caccianiga99,brinkmann00,landt01,caccianiga02_apj,beckmann03,padovani07}.   While these combined radio/X-ray selected samples include a significant fraction of IBLs in addition to LBLs and HBLs, they are of course sensitive to the varying flux limits and biases inherent to each multiwavelength survey.  

One practical limitation to recovering a large uniform sample from imaging surveys as described above is performing the necessary optical spectroscopic follow-up for (easily) thousands of candidates, among which only $\lesssim 10^2$ are typically actual BL Lacs.   It has been noted that BL Lacs tend to have characteristic X-ray-to-optical and optical-to-radio flux ratios differentiating them (at least statistically) from other radio/X-ray sources \citep[e.g.,][]{stocke89,stocke91,nass96}.  This feature has been exploited to increase the efficiency of BL Lac search algorithms by targeting radio/X-ray sources with broad-band flux ratios indicative of known BL Lacs.  Notably, the {\it Einstein} Slew Survey \citep{perlman96} and the Sedentary Multifrequency Survey \citep{giommi99,giommi05AA, piranomonte07} adopted this approach to recover large uniform  BL Lac samples ($\sim$60~in the case of the {\it Einstein} survey and $\sim$150~in the final Sedentary sample.)   While such broad-band color selection approaches can recover LBLs, IBLs and HBLs, it has been noted that the broad-band SED of BL Lacs can vary largely from object to object, and this selection approach is prone to missing some BL Lac subpopulations potentially in large numbers \citep{laurent97_thesis,laurent99}.  It should be noted however that the {\it Einstein} Slew Survey produced the first {\it large} BL Lac sample containing a sizable number of LBLs, IBLs {\it and} HBLs from a single selection technique; the Sedentary sample intentionally only attempted to recover HBLs, and it nicely complements other surveys more sensitive to LBLs \citep[particularly the Deep X-ray Radio Blazar Survey;][]{perlman98,landt01,padovani07}

The recent advent of digital large scale spectroscopic optical surveys has eased the difficulties of optical follow-up incurred by previous BL Lac surveys.  A purely optically selected sample was assembled from the 2df and 6df quasar surveys \citep{londish07}, and \citet{collinge05} additionally recovered hundreds of optically spectroscopically selected BL Lac candidates from the Sloan Digital Sky Survey \citep[SDSS,][]{york00}.  These optically selected samples do not entirely rule out the existence of radio-quiet BL Lacs; however, if such a population exists they must be extremely rare \citep[e.g., see][]{stocke90}.  

The $\sim$10$^3$~currently known BL Lacs were derived  from many different surveys, each with slightly different selection criteria and biases.   In this paper, we present a \nsamp\ object radio sample with homogeneous data coverage selected jointly from  the SDSS and from the Faint Images of the Radio Sky at Twenty-Centimeters \citep[FIRST,][]{becker95} survey.  This sample is one of the largest BL Lac samples yet assembled.  We also recently published in \citet{anderson07} a 266 object X-ray sample selected jointly from SDSS spectroscopy and from the Rosat All Sky Survey \citep[RASS,][]{voges99,voges00}.  Our X-ray sample recovered LBLs, IBLs and HBLs, and a post-selection correlation with FIRST and the NRAO VLA Sky Survey (NVSS) revealed only 7 objects currently lacking radio detections (possibly, but not certainly, because their radio emission lies below the FIRST/NVSS catalog flux limits.)  Both our separate SDSS/FIRST and SDSS/RASS samples are large with similar systematics given their similar recovery methods in SDSS.  These samples therefore provide an interesting opportunity for comparing radio and X-ray selected BL Lacs in a uniform manner.  The sheer size of our SDSS/FIRST sample, when supplemented by a rigorous understanding of selection biases among the different wavebands, may eventually permit application to a range of BL Lac topics, from AGN unification, to host galaxy properties, to BL Lac evolution.  

Our purpose here is to present our radio selected sample with some preliminary analysis.  We defer detailed investigations and comparisons with other samples to a later paper.   In \S \ref{sec:sampsel} we present our selection criteria.  Potential contaminants to our sample are discussed in \S \ref{sec:contaminants}, and our catalog is given in \S \ref{sec:catalog}.   Some preliminary results and future applications of our sample are discussed in \S \ref{sec:disc}, with a summary in \S \ref{sec:summary}.  Throughout we use $H_0= 71$~km s$^{-1}$~Mpc$^{-1}$,~$\Omega_m=0.27$, and $\Lambda_0=0.73$.

%%%%%%%%%%%%%%%%%%%%%%%%%%%%%%%%
%%%%%%%%%%%%%%%%%%%%%%%%%%%%%%%%
\section{Sample Selection}
\label{sec:sampsel}
We select BL Lac candidates from the combination of the SDSS Data Release 5 (DR5) spectroscopic database \citep{adelman07} in the optical, and from the FIRST survey  catalog \citep{becker95} in the radio.   The SDSS is a multi-institutional effort to image $10^4$~deg$^2$~of the north galactic cap in 5 optical filters covering 3800 to 10,000 \AA\ \citep[e.g., see][]{fukugita96, gunn98}, with follow-up moderate resolution spectroscopy ($\lambda/\Delta \lambda$$\sim$$1800$) of 10$^6$~galaxies, 10$^5$~QSOs, and 10$^5$~unusual stars \citep[e.g., see][]{york00}.  Data is taken with a special purpose 2.5-meter telescope located at Apache Point Observatory \citep[see][]{gunn06}, with astrometric accuracy at the $\sim$100 milli-arcsec level at the survey limit of $r\sim22$\ \citep{pier03} and typical photometric precision of approximately 0.02-0.03 mag \citep{ivezic04}; 640 simultaneous spectra are obtained over a 7 deg$^2$~field with a multifiber optical spectrograph.  Further technical details can be found in \citet{stoughton02}.  The SDSS DR5 spectroscopic database contains spectra for $\sim$$10^6$~objects over $\sim$$ 5700$~deg$^2$, and the DR5 Quasar Catalog \citep{schneider07}\ contains 77,429 quasars with reliable spectroscopic redshifts.  Our selection process identifies a very small fraction of objects ($<0.05\%$) in the DR5 Quasar Catalog as BL Lacs (see \S\ref{sec:catalog}).

The FIRST radio survey imaged approximately the same 10$^4$~deg$^2$~of the North Galactic Cap in the VLA's B configuration (5$''$ resolution) at 1.4 GHz.  Three-minute snapshots were co-added together, and sources were extracted via specialized algorithms \citep[see][]{white97} with a catalog flux limit of 1 mJy  and positional uncertainties better than $1''$.  The FIRST survey is by design well suited both in sky coverage and depth for matching optical counterparts in SDSS, and about one-third of FIRST sources have detectable optical counterparts in the SDSS photometric survey \citep{kimball08_subm}.  There are 37,456 objects in the SDSS DR5 spectroscopic database that are flagged as FIRST radio sources.

The target selection algorithms for follow-up SDSS spectroscopy are well suited for recovering BL Lacs.  Please refer to \citet{anderson03} for a thorough discussion.  Briefly,  one goal of the SDSS is to identify optical counterparts to $\sim$10$^4$\ RASS X-ray sources.   At least one source with $g,r$~or~$i<20.5$~within many $\sim$$1'$ RASS error circles is targeted for spectroscopy.  This X-ray counterpart algorithm targets many BL Lacs, as sources whose SDSS positions additionally match within 2$''$ to a FIRST source are given highest priority, followed by objects with quasar-like or otherwise strange SDSS colors.  The SDSS spectroscopic sample is even more advantageous for BL Lac recovery in that BL Lac spectra are additionally obtained through other quasar, serendipity, or galaxy spectroscopic target selection algorithms \citep[see][]{eisenstein01, richards02, strauss02}.

\begin{figure}
\centering
\includegraphics[scale=0.50]{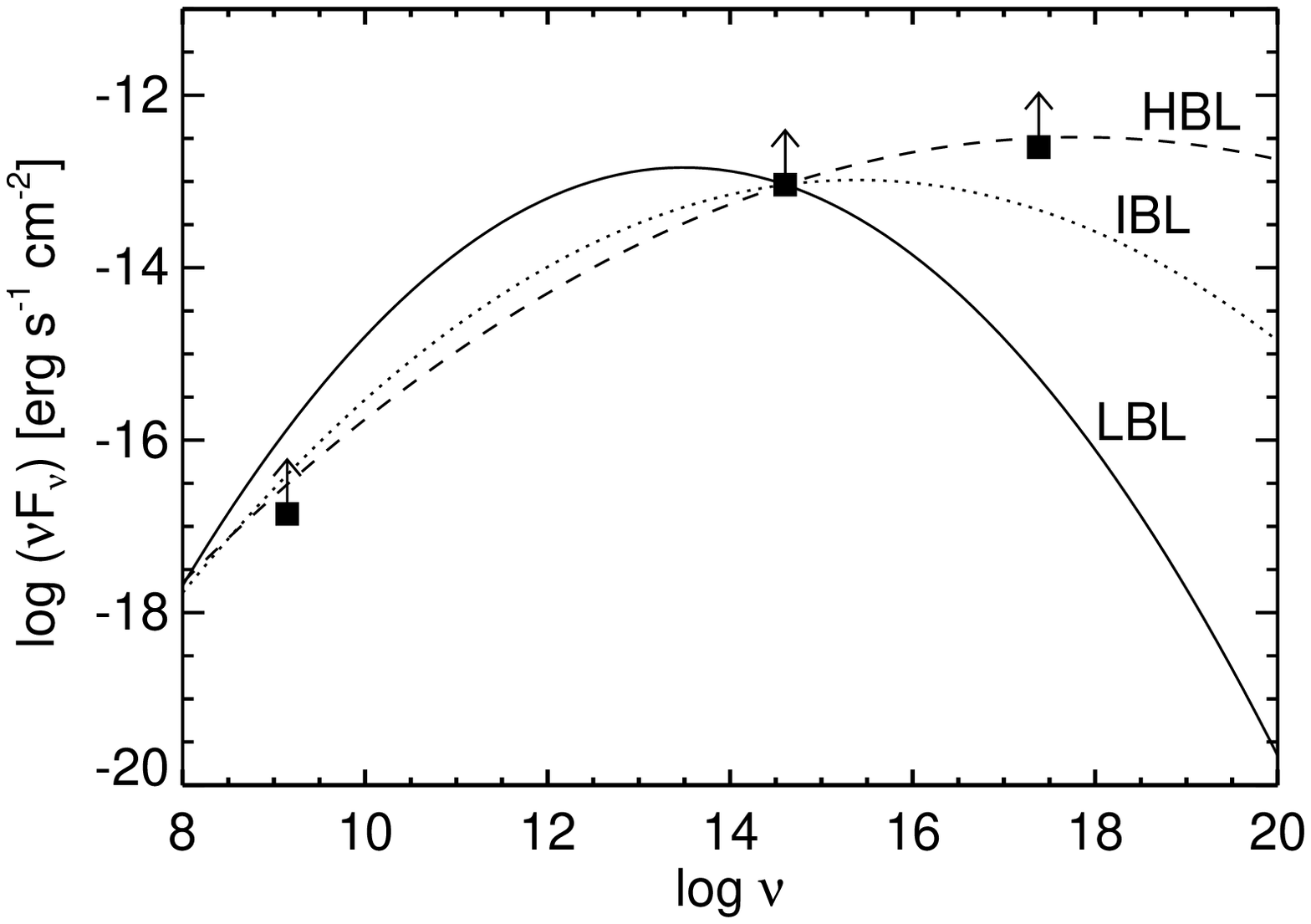}
\caption{
Average SEDs for LBLs ({\it solid line}), IBLs ({\it dotted line}), and HBLs ({\it dashed line}) in \citet{nieppola06}, normalized to $i= 20.5$~mag.  Approximate flux limits for FIRST in the radio (1 mJy at 1.4 GHz), SDSS spectroscopy in the optical ($i \lesssim 20.5$), and RASS in the X-ray ($\sim$2.5$\times$10$^{-13}$~erg~s$^{-1}$~cm$^{-2}$ from 0.1-2.4 keV, see \citealt{anderson07}) are shown as squares with arrows from left to right, respectively.  At the optical spectroscopic flux limit, we do not expect to miss many BL Lacs by imposing a radio flux limit, unless the SDSS object does not overlap with the FIRST survey's sky coverage, or if the object was in a low flux state during the FIRST observations.   Note that applying the RASS flux limit could bias our sample toward HBLs.  Although we do not formally impose an X-ray flux limit to our sample, the RASS flux limit is implicit in the SDSS spectroscopic target algorithms.  The RASS flux limit therefore could affect the makeup of our sample (see \S\ref{sec:disc}).  Our conclusion remains unchanged if we choose to normalize each SED to $g$\ or $r= 20.5$.
}
\label{fig:sed}
\end{figure}

Although one might initially expect that our emphasis on radio detection for inclusion would bias our sample toward  excess LBLs, we do not believe this is an important concern.  It is true that by requiring a detection in FIRST we are prone to missing some BL Lacs.  For example, some regions of the sky might not have data coverage in FIRST; and,  some BL Lacs with SDSS spectroscopy might have been in a low flux state during the FIRST observations and would therefore be missed; also, we are not highly sensitive to the radio-weak/quiet tail of the BL Lac population.  While very interesting, radio-weak/quiet BL Lacs must be extremely rare if they exist \citep[e.g., see][]{stocke90,londish07}; omission of a few would not affect our sample in a statistical sense.    However, most BL Lacs at the faint flux limit for SDSS spectroscopy ($g$, $r$, or $i < 20.5$) will have radio fluxes larger than the 1 mJy flux limit of the FIRST survey at 1.4 GHz.   Figure \ref{fig:sed} shows typical SEDs for an LBL, an IBL, and an HBL \citep[SEDs were kindly provided by Elina Nieppola, see][]{nieppola06} as a solid line, dotted line, and dashed line respectively; the SEDs are normalized to $i = 20.5$.  Overplotted as black squares with arrows are typical flux limits for FIRST, SDSS spectroscopy, and RASS from left to right, respectively.  
So, while it is likely there are some BL Lacs with SDSS spectroscopy not detected in FIRST, we do not appear to be significantly biasing our sample toward or against one particular BL Lac subclass by applying the radio flux limit to our selection criteria.  Note, however, that applying the RASS flux limit (taken from \citealt{anderson07}\ as 2.5$\times$10$^{-13}$~erg~s$^{-1}$~cm$^{-2}$\ from 0.1-2.4 keV) would bias our sample toward HBLs.  While we do not explicitly apply this flux limit, it does play a role in determining which objects are targeted for SDSS spectroscopy.  Therefore, objects in BL Lac samples derived from SDSS spectroscopy could be biased toward HBLs (see \S\ref{sec:disc} and also \citealt{collinge05}).  The relatively higher RASS flux limit could also explain why we recover almost twice as many BL Lac candidates from FIRST/SDSS as \citet{anderson07} did from RASS/SDSS over the same area of the sky.

%%%%Typical BL Lac Spectra (from Nieppola et al 2006)%%%%%%
%%%%%%%%%%%%%%%%%%%%%%%%%%%%%%%%

\subsection{Selection Criteria}
Following the process detailed below, we impose the following criteria for BL Lac selection: 1) the SDSS position matches within 2$''$~to a FIRST radio source; 2) SDSS spectroscopy shows no emission line with a rest-frame equivalent width ($REW$) larger than 5~\AA\ (see \S\ref{sec:maninspect}); and 3) the \ion{Ca}{2}~H/K~break (if present) has a measured depression $C\le0.40$.  The \ion{Ca}{2}~H/K~depression parameter $C$ describes the fractional change in continuum blueward and redward of 4000~\AA~\citep[e.g., see][]{landt02}:
$$ C=\frac{\langle f_{\nu,r}\rangle - \langle f_{\nu,b}\rangle}{\langle f_{\nu,r}\rangle}  =  0.14 + 0.86 \left ( \frac{\langle f_{\lambda,r}\rangle  - \langle f_{\lambda,b}\rangle}{\langle f_{\lambda,r}\rangle} \right ),$$
where $\langle f_{\nu,b}\rangle$~and $\langle f_{\nu,r}\rangle$~refer to the average fluxes per unit frequency just blueward (3750-3950 \AA) and just redward (4050-4250 \AA) of the H/K  break respectively.  The quantities $\langle f_{\lambda,b}\rangle$~and $\langle f_{\lambda,r}\rangle$~similarly refer to the average fluxes per unit wavelength.  These spectral constraints are commonly used in the recent literature, and they are consistent with, albeit more restrictive than, those introduced by \citet{marcha96}.  We also take additional precautions to reject objects that might formally pass these criteria but are not BL Lacs (e.g., stars, post-starburst galaxies, obvious broad absorption line quasars, etc.; see \S \ref{sec:contaminants}).

We note that our adopted selection criteria are empirically based and thus somewhat arbitrary  (e.g.,  see \citealt{rector99} and \citealt{caccianiga99} for relevant discussions pertaining to selection issues).      For example, there is no definitive \ion{Ca}{2}~H/K break value that serves as an obvious transition between beamed and unbeamed radio galaxies.   And, we could be missing some nearby low-luminosity BL Lacs, as described by the Browne-March{\~a} effect \citep{browne93, marcha95}: weakly beamed BL Lacs could be so dominated by starlight from their host galaxies that they are misidentified as radio galaxies or as clusters of galaxies.\footnote{Note that this issue is somewhat alleviated by using $C=0.40$\ as the H/K break cutoff instead of the more restrictive value of 0.25 used in some older samples.}  We discuss potential contamination from radio galaxies and clusters of galaxies in sections \ref{sec:radgal} and \ref{sec:clustgal} respectively.  Also, BL Lacs with significant host galaxy contamination can have $REW$'s that are measured to be larger than 5 \AA\ when referenced to their starlight contaminated continua instead of only their AGN continua \citep{marcha96}.  Our selection approach misses such objects.   Despite these well-studied issues, our adopted spectral criteria are also commonly applied in much of the recent literature.

%%%%%%%%%%%%%%%%%%%%%%%%%%
\subsection{Initial Query of the SDSS Database}
The SDSS pipeline's measurements of spectral features (e.g., redshifts, line equivalent widths, H/K breaks, etc.)\ are of course only useful if the pipeline estimates the correct redshift.  The pipeline redshifts are quite reliable for objects with strong spectral features (e.g., over 98\% of the pipeline redshifts were accurate for the objects in the DR5 Quasar Catalog), but one should not blindly trust the redshifts for such weak-featured objects as BL Lacs.  We therefore placed lenient restrictions on our initial database query to ensure high completeness at the cost of an initially large false positive rate.  We ran the following two-part query (followed by manual inspection of all SDSS spectra, see \S \ref{sec:maninspect}):

\begin{enumerate}
\item We return all SDSS spectra within 2$''$~of a FIRST source for which the pipeline has little confidence in the measured redshift.  This is achieved by selecting all spectra (that are not blank sky for calibration) that the database flags {\tt SPECZSTATUS} as {\tt `not\_measured'} or {\tt `failed'}.  This query returned 589 spectra.

\item For those objects with more confident pipeline redshifts (i.e., {\tt SPECZSTATUS}~$\neq$~{\tt `not\_measured'} or {\tt `failed'}), we return all spectra within 2$''$~to a FIRST source with the conservative constraints that (i) no emission line (of Ly$\alpha$, \ion{C}{4}, \ion{C}{3}], \ion{Mg}{2}, H$\beta$, [\ion{O}{2}], or H$\alpha$) has a pipeline measured $REW > 10$~\AA, and (ii) the H/K break (if present) has a pipeline measured $C<0.41$.  This query returned 1803 spectra.  Not every spectrum in this set has a reliable pipeline redshift.  However, as expected, essentially all spectra lacking reliable pipeline redshifts also have very weak spectral features, and our query correctly retains these objects.   Objects with spectral features too strong to be considered BL Lacs have correct pipeline redshifts in the vast majority of cases and are therefore rejected by the query as intended.  
 \end{enumerate}
 
 Out of $\sim$37,000 total FIRST matches in the DR5 spectroscopic database, our query yielded 2392 spectra as potential BL Lac candidates.  We do not consider X-ray information for selection.  However, if an SDSS source matches within $1'$\ to a RASS source, we also keep track of its X-ray properties listed in the DR5 database.
%%%%%%%%%%%%%%%%%%%%%%%%%%%
\subsubsection{Database Query Completeness and Efficiency}
\label{sec:queryeff}
We run two tests to check that our database query returns the vast majority of previously-known SDSS BL Lacs.  First, we compare our query results to \citet{collinge05}'s 183 probable SDSS optically selected BL Lac candidates that are additionally flagged as FIRST radio sources in their catalog.  \citeauthor{collinge05}~similarly require $C\leq 0.40$~and $REW \leq 5$~\AA, but their exact recovery method differs from ours.  172 of their 183 FIRST/SDSS objects appear in the DR5 database as FIRST sources and pass our signal-to-noise cut (see \S \ref{sec:signoise}).  Of these 172 sources, our database query fails to recover only a single spectrum.  We miss this spectrum because the pipeline measured a REW larger than 10~\AA~for an assumed \ion{Mg}{2} feature that is likely just noise.  
	
Second, we perform an identical query for X-ray selection by replacing the 2$''$~match to a FIRST source with a 1$'$~positional match to a RASS source.  We then compare results from this X-ray query to the \citet{anderson07} SDSS/RASS X-Ray selected AGN sample.  \citeauthor{anderson07}~queried the DR5 spectroscopic database for SDSS spectra that matched within 1$'$~to a RASS source, placing no additional constraints on the pipeline measured spectral feature  strengths or redshift flags.    Their manual inspection of the resulting 15,129 spectra yielded 266 X-ray selected BL Lac candidates (also defined by $REW \leq 5$~\AA~and $C \le 0.40$).  Our revamped X-ray query (i.e., including cuts on equivalent width and H/K break) weeds the $\sim$10$^4$~SDSS/RASS positional coincidences in DR5 to  $\sim$2000 objects; we recovered all but 3 spectra included in the \citeauthor{anderson07}~BL Lac sample.  The missed objects include a borderline BL Lac candidate with $C=0.4$~exactly (as measured by \citeauthor{anderson07}; the SDSS pipeline measured $C>0.41$), and two examples where the pipeline overestimated a single emission line's $REW$\ because the line lies in an abnormally noisy spectral regime.   

Although our database query is by design inefficient, we conclude that it is excellent in producing a manageable number of candidates while retaining essentially all previously-identified BL Lac objects  in the DR5 spectroscopic database.  

%%%%%%%%%%%%%%%%%%%%%%%%%%%%%%%
\subsection{SDSS Spectral Signal-to-Noise Cut}
\label{sec:signoise}
We apply a signal-to-noise cut to the SDSS spectra of potential BL Lacs, similar to \citet{collinge05}.  We calculate the signal to noise over three separate 500~\AA~wavelength regions centered on 4750, 6250, and 7750~\AA\ (each wavelength region typically contains approximately 450, 350 and 300 pixels respectively); these spectral regions are typically high-quality and fall within the $g,~r,$~and $i$~filters respectively.  Over each spectral region we calculate the signal (S) and the noise (N) such that $S/N = \left( \sum_i f_{\lambda,i}/\sigma_i^2 \right ) / \left (\sum_i 1/\sigma_i^2 \right )^{1/2}$, where $f_{\lambda,i}$~and $\sigma_i$~represent the flux density and the estimated uncertainty in each spectral element $i$~respectively.  We retain all objects with $S/N > 100$\ in at least one of these spectral regions.  This constraint roughly corresponds to fiber magnitudes $g<20.5$, $r<20.3$, or $i<19.6$ for a typical spectroscopic plate, and it truncates our list from 2392 to 1560 objects. 

%$S=\sum_i \left (f_{\lambda,i}/\sigma_i^2 \right )/\sum_i \left (1/\sigma_i^2 \right )$ and the noise $N=\left ( \sum_i 1/\sigma_i^2 \right )^{-1/2}$, 

%%%%%%%%%%%%%%%%%%%%%%%%%%
\subsection{Manual Inspection of SDSS Spectra}
\label{sec:maninspect}
We manually inspect the remaining 1560 spectra, including visual verification that our $S/N$ cut did indeed only reject objects with poor signal to noise.   First, we flag each pipeline measured redshift as either reliable, as tentative, as a lower limit, or as bad.   We define reliable pipeline redshifts as those based on at least two spectral features at the same pipeline redshift.  Tentative pipeline redshifts are defined as those showing either only a single spectral feature, typically an emission line assumed to be \ion{Mg}{2}, or showing multiple weak yet potentially real spectral features at the same pipeline redshift.  For spectra only showing absorption features typically associated with intergalactic medium absorption lines (i.e., doublets like \ion{Mg}{2}~$\lambda$2796,2803, \ion{Fe}{2}~$\lambda$2374,2382, etc.), we assign redshifts equal to the lower limit derived from the absorption feature and mark those objects' redshifts as lower limits.  Note that these objects could have redshifts equal to the limit values.  Bad pipeline redshifts either show no spectral features, or the pipeline simply assigned an incorrect redshift (as judged in the visual verification process).  For objects with tentative and bad pipeline redshifts, we attempt to find a redshift that fits the spectrum better than the pipeline redshift.  If we find a better redshift (i.e., $\ge$2 spectral features at the same redshift,) then we change that spectrum's redshift status to reliable.   The small number of spectra that show only a single emission line are assigned tentative redshifts assuming the lone emission is \ion{Mg}{2}.  Remaining objects with bad pipeline redshifts are then reclassified as having unknown redshifts.

For spectra with reliable and tentative redshifts we measure the strongest emission line's $REW$~with {\tt IRAF}\footnote{IRAF is distributed by the National Optical Astronomy Observatories,
    which are operated by the Association of Universities for Research
    in Astronomy, Inc., under cooperative agreement with the National
    Science Foundation.}, retaining all objects for which this line's $REW \le 5$~\AA\ (assuming tentative redshifts are correct.)   A handful of objects with redshift lower limits from intervening absorption show a single emission feature; we measure this emission feature's $REW$\ assuming the limit value is the correct redshift.  We fit any blended lines with multiple Gaussian components as needed.   We do however reject candidates with H$\alpha$/[\ion{N}{2}] complexes that show broad H$\alpha$, even if our deblended line measurements show components with $REW \le  5$~\AA, under the assumption that we underestimate the H$\alpha$~line flux and overestimate the [\ion{N}{2}] line flux.  Most objects with tentative redshifts have such weak features that they undeniably pass the equivalent width cut, especially since they tend to have {\it observed-frame} equivalent widths $\le 5$~\AA.   Exceptions might be the small number of spectra that show only a single emission line.  For these objects, we estimate {\it rest-frame} $EW$~by assuming the lone emission is \ion{Mg}{2}~$\lambda 2800$. Objects with adequate $S/N$\ but unknown redshifts pass our equivalent width cut almost by definition.  For objects with reliable, tentative, or lower limit redshifts, we remeasure the \ion{Ca}{2}~H/K~break  parameter $C$~(if present) through an automated process, rejecting all objects with measured $C>0.40$ (assuming tentative redshifts and derived lower limits are correct.)  We also examine the SDSS finding charts of each surviving object to ensure none are obviously blended or contaminated by a nearby object. 

\begin{figure}
\centering
\includegraphics[scale=0.50]{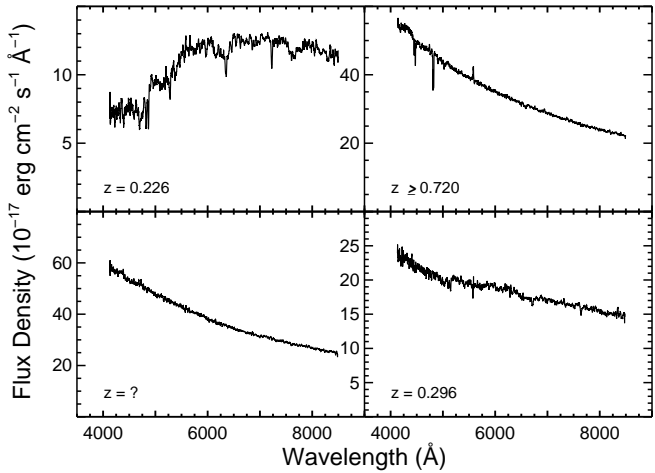}
\caption{
Four sample spectra of higher confidence BL Lac candidates, boxcar smoothed with a width of 9.  All 4 objects are additionally X-ray emitters.  
{\it Top left:} SDSS~J082814.20$+$415351.9, an example candidate with a reliable redshift ($z=0.226$) and prominent host galaxy contamination  (\ion{Ca}{2} H/K depression $C=0.333$).  
{\it Top right:} SDSS~J111757.24$+$535554.9, an example of a nearly featureless object with a lower limit redshift ($z\geq0.720$) derived from absorption doublets.  Here, absorption lines from \ion{Fe}{2} $\lambda 2344,2374,2383,2587,2600$\ and \ion{Mg}{2} $\lambda 2796,2804$\ are all  observed at the redshift quoted above.  Most objects in our sample with redshifts derived from absorption doublets typically only show a single absorption feature. 
{\it Bottom left:} SDSS~J115034.75$+$415440.1, an example of a featureless object with unknown redshift.  
{\it Bottom right:} SDSS~J120412.11$+$114555.4, an example of a candidate with a reliable redshift ($z=0.296$) and moderate host galaxy contamination ($C=0.131$).
We are careful not to reject objects based on weak features near night sky lines such as 5577 \AA.
}
\label{fig:smpbl}
\end{figure}

\begin{figure}
\centering
\includegraphics[scale=0.50]{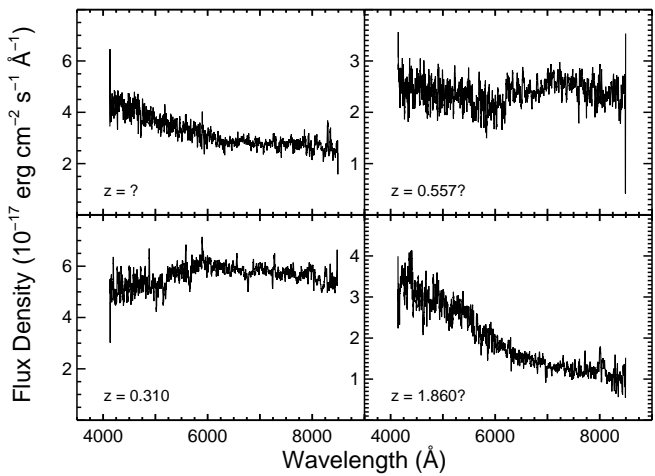}
\caption{
Four sample spectra of lower confidence BL Lac candidates, boxcar smoothed with a width of 9. 
{\it Top left:} SDSS~J082801.14$+$231217.6, classified as `BL?' due to low signal-to-noise.  This source has an unknown redshift and does not match to an X-ray source.
{\it Top right:} SDSS~J095507.88$+$355100.8, potentially shows moderate host galaxy contamination ($C=0.231$) at a tentative redshift $z=0.557$.  This object is classified as `BL?' due to low signal-to-noise, and it has an X-ray match in RASS.
{\it Bottom left:} SDSS~J132558.94$+$411500.3, shows moderate host galaxy contamination ($C=0.223$) at a reliable redshift $z=0.310$.  This source is classified as `BL?'\ because a broad H$\alpha$\ feature can have $REW$\ larger or smaller than 5 \AA\ depending on the continuum level chosen near the feature.  This source has an X-ray match in RASS.
{\it Bottom right:} SDSS~J165329.09$+$370511.4, classified as `BL?' due to low signal-to-noise.  This source has a tentative redshift ($z=1.860$) and does not have an X-ray match in RASS.
}
\label{fig:smpmaybe}
\end{figure}

Finally, we classify each remaining object as either a higher confidence BL Lac candidate (`BL') or as a lower confidence BL Lac candidate (`BL?').  We require objects classified as `BL' to have slightly higher quality spectra, that is $S/N>125$~in at least one of the three spectral regions described in \S \ref{sec:signoise}.  Objects can be classified as `BL?' for any of the following reasons:  (1) the largest calculated $S/N$, as described in \S \ref{sec:signoise}, has $100<S/N\leq125$;  (2) the spectrum passes all criteria, but the object appears close enough to another bright object in its SDSS finding chart that flux contamination of the fiber spectrum might be a concern; or  (3) it is unclear where to define the continuum near an emission line, and the line's $REW$~is measured as larger or smaller than 5~\AA~depending on the continuum choice.  We classify \nbl\  objects as `BL' and \nmaybe\  as `BL?'.  Some sample spectra for objects classified as `BL' and `BL?' are shown in Figures \ref{fig:smpbl} and \ref{fig:smpmaybe} respectively.

%%%%%%%%%%%%%%%%%%%%%%%%%%%%%%%%%%%%
%%%%Sample BL Lac Spectra%%%%%%

%%%%Sample BL? Spectra%%%%%
%%%%%%%%%%%%%%%%%%%%%%%%%%%%%%%%%%
%%%%%%%%%%%%%%%%%%%%%%%%%%%%%%%%%%

%%%%%%%%%%%%%%%%%%%%%%%%%%%
\subsection{Luminosity and Broad-Band Spectral Index Calculations}
\label{sec:fluxcalc}
We estimate rest-frame monochromatic luminosities per unit frequency in the radio, optical and X-ray at 5 GHz, 5000 \AA, and 1 keV, respectively.   We adopt a radio spectral index $\alpha_r=-0.27$, the average of the 1 Jy radio sample \citep{stickel91}; and we assume $\alpha_o=1.5$~in the optical (typical of \citealt{collinge05}'s probable BL Lac candidates) and $\alpha_x=1.25$~in the X-ray \citep[a value approximately intermediate between LBLs and HBLs,][]{sambruna96,sambruna97}.    All spectral indices are defined as $f_\nu\sim\nu^{-\alpha_\nu}$.  We choose the above reference frequencies and spectral indices for consistency with the bulk of the literature.  

Radio flux densities at 1.4 GHz are taken from the integrated FIRST flux densities in the SDSS database.  The SDSS filter that includes 5000~\AA\ in each object's rest frame is used to estimate optical flux densities.  We assume tentative and lower limit redshifts are correct, and we use the $r$~filter for spectra with unknown redshifts (as 5000 \AA~would fall in the $r$~filter at the median redshift of our sample.)  Extinction corrected magnitudes (using the \citealt{schlegel98} dust maps) are converted to flux densities using a zero-point of 3631 Jy and effective wavelengths 4686, 6165, 7481, and 8931~\AA~for $g$, $r$, $i$, and $z$~respectively.  We use point spread function (psf) magnitudes to reduce potential host galaxy contamination in our optical luminosity calculations, but we note that we might still be overestimating optical luminosities for some nuclei.

For the X-ray, we use the \citet{stark92} \ion{H}{1} maps and the {\tt colden} tool in CIAO \citep{ciaoref} to estimate the hydrogen column density along each sightline.  We then use these column densities and the Portable, Interactive Multi-Mission Simulator \citep[PIMMS,][]{pimmsref} to convert ROSAT PSPC X-ray count rates to absorbed and unabsorbed broad-band (0.1-2.4 keV) fluxes.  For each spectrum lacking an X-ray detection we estimate its RASS exposure time as the exposure time for the X-ray source in the RASS catalog closest to the SDSS source's position.  Upper limits on RASS count rates are then set as 6 counts \citep[the limit for inclusion in the RASS faint source catalog,][]{voges00} divided by the exposure time, and limits on X-ray fluxes are calculated as described above.   

Monochromatic luminosities are then calculated from the above radio and optical flux densities and the unabsorbed X-ray fluxes, assigning a redshift z=0.297 (the median reliable redshift of our \nbl\ higher confidence candidates) to objects with unknown redshifts.  We then calculate the broad-band spectral indices $\alpha_{ro}$, $\alpha_{ox}$, and $\alpha_{rx}$.\footnote{The broad-band spectral index, for $\nu_2>\nu_1$, is defined as $\alpha_{\nu_1\nu_2}=-\log(L_{\nu_2}/L_{\nu_1})/\log(\nu_2/\nu_1)$.  Here, $\alpha_{ro}=-\log(L_o/L_r)/5.08$, $\alpha_{ox}=-\log(L_x/L_o)/2.60$, and $\alpha_{rx}=-\log(L_x/L_r)/7.68$, where $L_r$, $L_o$, and $L_x$~are the specific luminosities (per unit frequency) at 5 GHz, 5000 \AA, and 1 keV respectively.}  Throughout, we make no attempt to correct luminosities for the effects of relativistic beaming.

%%%%%%%%%%%%%%%%%%%%%%%%%%
%%%%%%%%%%%%%%%%%%%%%%%%%
\section{Possible Sources of Contaminants}
\label{sec:contaminants}

%%%%%%%%%%%%%%%%%%%%%%%
\subsection{Positional Coincidences}
\begin{figure}
\centering
\includegraphics[scale=0.45]{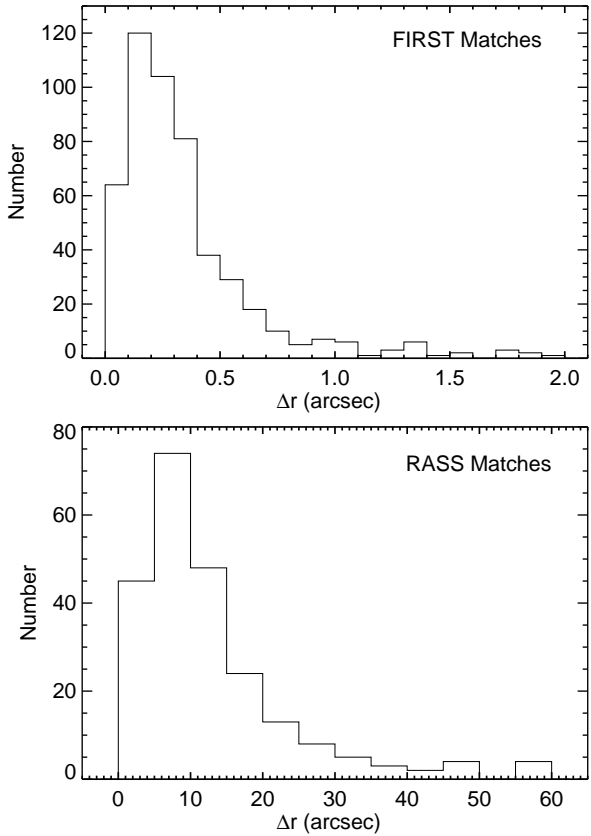}
\caption{{\it Top panel:} differences between SDSS and FIRST source positions for all \nsamp\ BL Lac candidates;  95\% of objects have SDSS postitions within $1''$\ of each FIRST position.  
{\it Bottom panel:} differences between SDSS and RASS source positions for the \nsampxray\ candidates with X-ray detections; 89\% of objects with X-ray detections have SDSS positions within $25''$\ of each RASS position.  These percentages are consistent with little contamination from positional mismatches between FIRST, SDSS, and RASS.}
\label{fig:dist}
\end{figure}

We have very little contamination from random mismatches between SDSS and FIRST source positions.  Figure \ref{fig:dist}~(top) shows differences between SDSS and FIRST positions for our \nsamp\  BL Lac candidates.  \citet{becker95} find FIRST positions to be accurate to $<1''$ at the 90\% confidence level, and $95\%$\ of our SDSS/FIRST BL Lac candidates have SDSS source positions within $1''$\ of the FIRST radio position.  While X-ray emission is not a criterion for inclusion in our sample, the bottom panel of Figure~\ref{fig:dist} pleasantly confirms that we have a low false X-ray match rate as well:   of our \nsampxray\ X-ray matches, 89\% have positional differences less than $25''$.  This is consistent with expectations from \citet{voges99}, who find 90\% of objects in the RASS catalog with stellar optical counterparts to have X-ray source positions within $25''$\ of the optical counterparts' positions. 

As another test on the level of contamination, we recorrelate all \nsamp\  BL Lac candidates to the FIRST and RASS catalogs after randomly offsetting the SDSS source positions by up to $\pm20''$\ and $\pm10'$\ respectively (i.e., 10 times the corresponding match radii.)  We perform the recorrelation 10 times, requiring each offset to be larger than each match radius.    On average we obtain only 6 random matches to a FIRST or to a RASS source in each offsetting test, reaffirming a very low level of potential contamination ($\sim1\%$) due to positional mismatches. 

%%%%%%%%%%%%%%%%%%%%%%%%%%%%%%%%%%
%%%%Positional Matches between SDSS and FIRST/RASS%%%%
%%%%%%%%%%%%%%%%%%%%%%%%%%%%%%
%%%%%%%%%%%%%%%%%%%%%%%%%%%%%%

%%%%%%%%%%%%%%%%%%%%%%
\subsection{Radio Stars}
\label{sec:radiostars}
\begin{figure}
\centering
\includegraphics[scale=0.5]{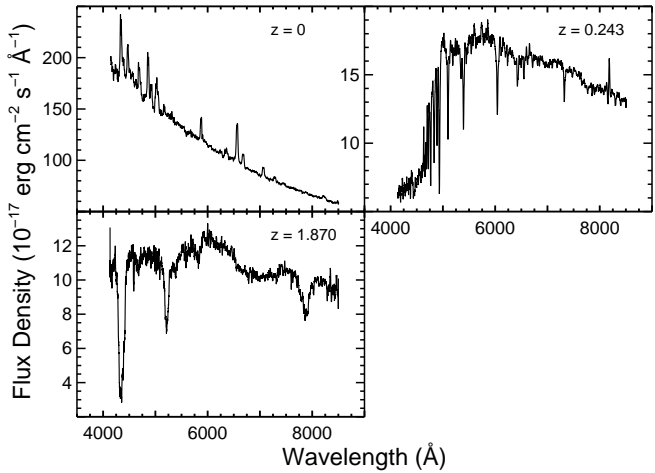}
\caption{
Sample SDSS spectra of objects rejected during manual inspection, boxcar smoothed with a width of 9.  
{\it Top left:}  FIRST~J102347.6$+$003841 (SDSS~J102347.68$+$003841.1), a previously known cataclysmic variable (CV, see \S \ref{sec:radiostars}).  This is the only CV recovered by our database query, and it was rejected during manual inspection based on emission line strength.
{\it Top right:} SDSS~J095842.64$+$631845.5 ($z=0.243$), an E+A galaxy (see \S \ref{sec:radgal}).  This object is rejected as a post star forming galaxy based on H$\delta$\ absorption.
{\it Bottom left:} SDSS~J123525.15$+$351242.6  ($z=1.870$).  This object is rejected as a broad absorption line quasar (BALQSO, see \S \ref{sec:bal}). 
}
\label{fig:smpcontam}
\end{figure}

Since our database queries do not include cuts on absorption features other than \ion{Ca}{2} H/K break strength, we need to remove potential stellar contaminants during manual inspection of the spectra.  Manual inspection revealed hundreds of radio stars of various spectral classes identified via common stellar features consistent with zero redshift.  Most of these objects would formally fulfill our BL Lac selection criteria, but they are rejected since they clearly possess stellar spectra.    We defer substantive discussion of the radio emitting stars for a later publication.   As an example however, we show the spectrum of a previously known cataclysmic variable \citep[FIRST~J102347.6$+$003841,][]{bond02} in the top left panel of Figure \ref{fig:smpcontam} that was rejected during manual inspection based on emission line strength.   

\begin{figure}
\centering
\includegraphics[scale=0.45]{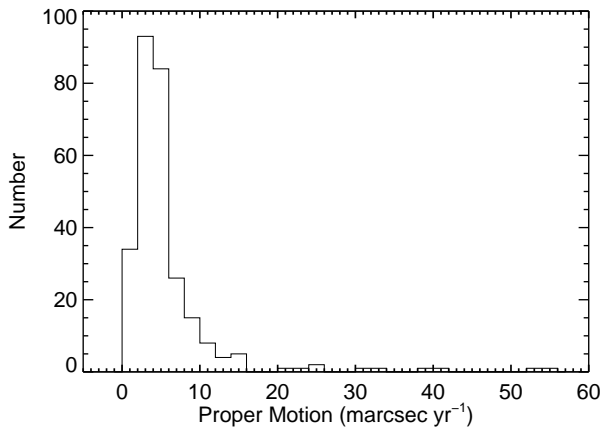}
\caption{Proper motions for 279 BL Lac candidates with $r<20$; these 279 candidates all have proper motion information available in the SDSS database, and 91\% show $\mu< 10.6$\ milli-arcsec year$^{-1}$, consistent with little stellar contamination in our final sample.  About half of the objects with $\mu>10.6$\ milli-arcsec year$^{-1}$\ have reliable or tentative redshifts.}
\label{fig:pm}
\end{figure}

To test the potential level of stellar contamination in our final sample, we examine proper motion information from the SDSS+USNO-B proper motion catalog \citep{munn04}.  Proper motions are available for $\sim$60\% of our sample, and we assume objects with proper motion information are representative of the rest of our sample.   \citet{munn04} find 95\% of a sub-sample of spectroscopically confirmed quasars with $r<20$~in the SDSS+USNO-B catalog to have proper motions $\mu < 10.6$\ milli-arcsec~year$^{-1}$.  Figure \ref{fig:pm} shows a histogram of proper motions for 279 BL Lac candidates with proper motion detections and $r < 20$;  $91\pm6\%$\  show $\mu<10.6$~milli-arcsec year$^{-1}$.  About half of our BL Lac candidates with measured $\mu>10.6$\ milli-arcsec year$^{-1}$\ have reliable or tentative redshifts.  We therefore do not believe galactic sources to be important contaminants to our final sample.   

We list 13 BL Lac candidates with large cataloged  proper motions ($\mu \ge 10.6$\ milli-arcsec yr$^{-1}$) and unknown redshifts in Table \ref{tab:bigpm}.  We give object names in column (1),  our classifications as a higher confidence (`BL') or lower confidence (`BL?') candidate in column (2), the measured proper motions (in milli-arcsec yr$^{-1}$) from the SDSS DR5 database in column (3), psf-magnitudes in the $r$\ filter in column (4), and integrated radio fluxes from FIRST in mJy in column (5).  Only one object (SDSS~J094432.33+573536.1) has an X-ray detection in RASS.  These sources could potentially be unusually featureless radio stars, and they deserve careful follow-up, such as, for example, proper motion measurements with very long baseline interferometry.  We do not reject these sources, however, since we do not have proper motion information for every SDSS object and the number of objects with large proper motion measurements is consistent with random expectations from \citet{munn04}.  

%%Large Proper Motions%%%
%\begin{turnpage}
\begin{deluxetable}{ccccc}
\tablewidth{0pt}
\tablecaption{Objects with $\mu \ge 10.6$\ milli-arcsec yr$^{-1}$\ and Unknown Redshifts \label{tab:bigpm}}
\tablehead{
	   \colhead{SDSS Name} %1
	& \colhead{Class.} %2
	& \colhead{$\mu$} %3
	& \colhead{$r$} %4
	& \colhead{Radio Flux} %5
\\
	    \colhead{(J2000)} %1
	 & %2
	 & \colhead{(marcsec/yr)} %3
	 & \colhead{(mag)} %4
	 & \colhead{(mJy)} %5
}	
\startdata
075547.72$+$432744.7 &  BL? &   11.7 &   19.76 &    2.63  \\
094432.33$+$573536.1 &  BL  &  52.5  &   19.87 &    1.86   \\
094441.47$+$555752.9 &  BL? &   54.5 &   19.89 &   24.67 \\
100040.67$+$531911.7 &  BL? &   13.9 &   20.36 &    2.02 \\
103845.88$+$392735.0 &   BL &   12.1 &   19.23 &    7.09 \\
113115.50$+$023450.2 &   BL &   22.4 &   18.53 &   13.29 \\
120355.34$+$581945.5 &   BL &   14.2 &   18.57 &   41.81 \\
125032.58$+$021632.1 &   BL &   14.1 &   19.21 &  468.49 \\
130146.31$+$441619.2 &   BL &   38.8 &   18.64 &   53.64 \\
135158.20$+$554210.8 &   BL &   40.6 &   19.22 &   83.40 \\
141208.23$+$383521.6 &  BL? &   13.3 &   20.16 &    3.35 \\
164301.06$+$322104.0 &   BL &   25.1 &   18.51 &   22.21 \\
165542.81$+$324420.0 &  BL? &   11.2 &   18.68 &   31.63 \\
\enddata
\end{deluxetable}
%\end{turnpage}

%%%%%%%%%%%%%%%%%%%%%%%
%%%%%Sample Contaminating Spectra%%%%%
%%%%%%%%%%%%%%%%%%%%%%%%%%%%%%%%%
%%%%%Proper Motion Histograms%%%%%
%%%%%%%%%%%%%%%%%%%%%%%%%
%%%%%%%%%%%%%%%%%%%%%%%%%

%%%%%%%%%%%%%%%%%%%%%%
\subsection{Radio Galaxies}
\label{sec:radgal}
Radio emitting galaxies are contaminants returned in relatively large numbers by our database query.  Most contaminating radio galaxies are easily identified through emission lines with $REW>5~$\AA\ and are rejected during manual inspection.  Some radio galaxies however can be more challenging to recognize.  For example, some objects show spectra with a weak H/K break due to an increased blue continuum not from nuclear activity but rather from a recent episode of star formation (e.g., E+A ellipticals or other post-starburst galaxies.)  Post-starburst galaxies with weak emission lines would formally pass our selection criteria.  We therefore reject post-starburst galaxies if H$\delta$~shows a $REW$~larger than 5~\AA~in absorption\footnote{H$\delta$~absorption is a common proxy for post-starbust galaxies \citep[e.g.,][]{goto03}.}, and/or if \ion{Ca}{2}~H+H$\epsilon$~is significantly stronger than \ion{Ca}{2}~K (as suggested by \citealt{stocke91} to reduce galaxy contamination in BL Lac samples).  A sample spectrum of an E+A galaxy that is rejected during manual inspection based on strong H$\delta$\ absorption is shown in the top right panel of Figure \ref{fig:smpcontam}.

A small fraction of spectra might show small H/K breaks that are not weakened by either nuclear activity or star formation, but they could rather be passive elliptical galaxies with intrinsically small H/K breaks; this does not appear to be a strong source of contamination however, as \citet{marcha96} point out that the \citet{dressler87} sample of over 700 early type galaxies has a mean H/K break of $C=0.49$~and less than 5\%~show $C<0.40$.   While requiring the more stringent (and historical) criterion of $C\le0.25$\ for inclusion in BL Lac samples would essentially eliminate any concern of contamination from passive elliptical galaxies \citep[e.g., see][]{stocke91}, \citet{marcha96} argue that doing so misses some weakly beamed BL Lacs.   There appears to be a smooth transition between BL Lac and radio galaxy spectra, and as noted in \S \ref{sec:sampsel}, the exact \ion{Ca}{2}~H/K break cutoff used to define BL Lacs is thus somewhat arbitrary.   BL Lac selection criteria are commonly extended to include objects with $C\le0.40$, as this H/K break cutoff would include most beamed AGN, but it is not large enough to produce significant contamination from elliptical galaxies with intrinsically small H/K breaks.   We therefore follow the bulk of the recent literature and use $C\le0.40$, acknowledging that this might allow a small number of contaminating passive elliptical galaxies into our sample.  Polarization and variability follow-on of objects with $0.25<C\le0.40$\ would be helpful for confirming their classification as BL Lacs.

%%%%%%%%%%%%%%%%%%
\subsection{Clusters of Galaxies}
\label{sec:clustgal}
See \citet{rector99} for a discussion on recognizing BL Lacs from clusters of galaxies.  We investigate in a preliminary fashion potential contamination from non-BL Lac objects residing in nearby clusters of galaxies by correlating our BL Lac candidates to the \citet{koester07} SDSS maxBCG cluster catalog.   The \citeauthor{koester07}\ catalog contains 13,823 photometrically selected clusters with velocity dispersions $\gtrsim400$~km s$^{-1}$~from $z=0.1-0.3$~over 7500 deg$^2$.  Photometric redshifts are given for each cluster, which are estimated to be accurate to $\Delta z \approx 0.01$~from comparison to available spectroscopic redshifts.  We perform the matching with a fixed projected physical linear aperture of radius 0.5 Mpc at the photometric redshift of each cluster.  Among our BL Lacs, 12 higher confidence candidates and 1 lower confidence candidate match to a maxBCG cluster.  Of these 13 matches, 4 of the BL Lac candidates have no reliable redshift, 5 have reliable redshifts within $\Delta z\sim0.01$~of the cluster's photometric redshift, and 4 have reliable, tentative or lower limit redshifts larger than the cluster's photometric redshift by at least 0.01.  Table \ref{tab:clust} gives the SDSS name for these 13 BL Lac candidates, our classification (`BL' or `BL?'), the angular separation between the BL Lac position and the brightest cluster galaxy, each BL Lac candidate's spectroscopic redshift, and each cluster's photometric redshift.   All objects pass our selection criteria, even the more stringent constraints aimed toward rejecting post star forming radio galaxies (see \S \ref{sec:radgal}); only 3 objects have measured \ion{Ca}{2}~H/K break depressions $C>0.25$~(the largest  is 0.35).  Thus, contamination from post star forming galaxies or radio galaxies with weak H/K breaks residing in clusters is likely small, and we retain all 13 objects in our final sample.  BL Lacs tend to avoid rich clusters at low redshifts \citep[$z\lesssim 0.35$, see][]{wurtz97}, and it is therefore reassuring that only a very small number of BL Lac candidates were returned by our cluster correlation.  

We estimate the expected number of chance superpositions by randomly offsetting our BL Lac positions by $\pm 5-50'$\ (the largest angular radius used in our original correlation was $4.5'$) and recorrelating to the cluster catalog 10 times; we find 5 chance superpositions on average, and it is reasonable that the 8 BL Lac candidates in Table \ref{tab:clust}\ with unknown redshifts or redshifts very different than each cluster's redshift could be chance superpositions.  Some of the 13 matches (especially those with a BL Lac redshift similar to a cluster redshift) might be examples of clusters of galaxies hosting a BL Lac. 

%%%%%%%%%%%%%
%%Cluster Matches%%%
\begin{deluxetable}{ccccc}
\tablewidth{0pt}
\tablecaption{Matches to maxBCG Clusters\label{tab:clust}}
\tablehead{
	   \colhead{SDSS Name} %1
	& \colhead{Class.} %2
	& \colhead{Sep.} %3
	& \colhead{BL Lac} %4
	& \colhead{Cluster} %5
\\
	    \colhead{(J2000)} %1
	 & %2
	 & \colhead{(arcsec)} %3
	 & \colhead{$z_{\rm spec}$} %4
	 & \colhead{$z_{\rm photo}$} %5
}	
\startdata
082814.20$+$415351.9 & BL  & 27  & 0.226 & 0.224 \\
091848.58$+$021321.8 & BL  & 82  & \nodata & 0.278 \\
092642.18$+$080300.2 & BL? & 186 & \nodata & 0.116 \\
101504.13$+$492600.6 & BL  & 126 & \nodata & 0.151 \\
113118.63$+$580858.8 & BL  & 147 & 0.360 & 0.167 \\
122008.29$+$343121.7 & BL  & 64  & 0.870\tablenotemark{a} & 0.246 \\
124700.72$+$442318.8 & BL  & 63  & 1.812 & 0.197 \\
134105.10$+$395945.4 & BL  & 128 & 0.172 & 0.176 \\
144248.28$+$120040.2 & BL  & 52  & 0.163 & 0.159 \\
161541.21$+$471111.7 & BL  & 68  & 0.199 & 0.205 \\
161823.58$+$363201.7 & BL  & 173 & 0.734\tablenotemark{b} & 0.159 \\
162259.26$+$440142.9 & BL  & 210 & \nodata & 0.122 \\
163726.66$+$454749.0 & BL  & 73  & 0.192 & 0.203 \\
\enddata
\tablenotetext{a}{redshift is a lower limit}
\tablenotetext{b}{tentative redshift}
\end{deluxetable}
%%%%%%%%%%%%%%%%%
%%%%%%%%%%%%%%%%%%

%%%%%%%%%%%%%%%%%%%%%%%%
\subsection{$z>1$~BL Lac Candidates and Weak-Lined Quasars}
\label{sec:qso}
A follow-on polarization study of a subset of 42 of the \citet{collinge05} optically-selected BL Lac candidates revealed all nine  $z>1$~objects in that subset lack strong optical polarization (e.g., below 3\%), thereby raising doubts as to whether they should be classifed as BL Lacs \citep{smith07}. We note that these results might not be entirely relevant to our sample, as only two of their nine $z>1$\ objects with follow-on polarization observations were radio sources. However, SDSS has revealed a number of similar objects at very high redshift, often termed weak-lined quasars, whose optical spectra lack strong emission (like BL Lacs), but which also display low polarization and a range of radio emission extending to quite low values (unlike BL Lacs); see \citet{fan99}, \citet{anderson01}, and \citet{shemmer06}\ for discussions on SDSS weak-lined quasars. 

Thirty-one of our higher confidence BL Lac candidates and 14 of our lower confidence BL Lac candidates have $z>1$ (assuming tentative redshifts are correct.)  Of our 31 higher confidence BL Lac candidates with reliable or tentative redshifts at $z>1$, there are 5 that appear relatively radio weak ($\alpha_{ro}<0.2$), suggesting that some of our high-$z$~candidates are indeed atypical of BL Lacs.  Given our lack of polarization information we retain all $z>1$ objects.  Polarization, variability, and multiwavelength investigations would be enlightening in confirming their classification.

%%%%%%%%%%%%%%%%%%%%%%%
\subsection{Broad Absorption Line Quasars}
\label{sec:bal}
We also recover several objects that formally fulfill our BL Lac criteria but are rejected upon manual inspection of their spectra as obvious broad absorption line quasars (BALQSOs, see bottom left panel of Figure \ref{fig:smpcontam}).

%%%%%%%%%%%%%%%%%
\subsection{An Unusual Class of Radio Sources}
\label{sec:oddbals}
\citet{hall02} discovered two objects, SDSS J010540.75-003313.9 ($z$$=$1.179) and SDSS J220445.26+003141.9 ($z$$=$1.353), that show a very strange continuum drop-off blueward of a \ion{Mg}{2} absorption doublet; both are FIRST radio objects and neither are RASS X-ray sources.  Both objects also show broad H$\alpha$ in the near-infrared (Hall et al.\ 2008, in preparation).    \citeauthor{hall02}~were aware of only 2 other objects previously reported in the literature with similar spectra, FBQS~1503+2330 ($z$$=$0.492) and FBQS~1055+3124 ($z$$=$0.404), both of which are also radio sources \citep{white00}.  Initially, no entirely satisfactory explanation was found for these unusual objects  (see \citeauthor{hall02}'s \S6.1).  However, one promising possible explanation suggested by \citeauthor{hall02}\ is that they are extraordinarily unusual BALQSOs.  

\begin{figure}
\centering
\includegraphics[scale=0.48]{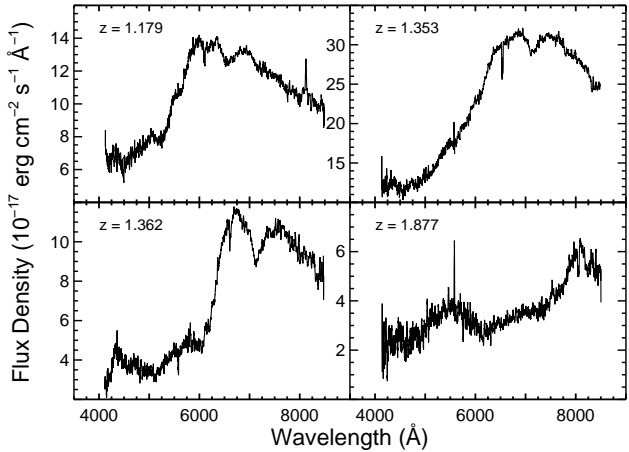}
\caption{Spectra of mysterious radio AGN \citep[possibly highly unusual BALQSOs;][]{hall02} as described in \S \ref{sec:oddbals}.  Spectra are ordered by increasing redshift and boxcar smoothed with a width of 9.  None match to a RASS X-ray source.
{\it Top left:} SDSS~J010540.75$-$003313.9 ($z=1.179$), published in \citeauthor{hall02}  This object formally passes our selection criteria, and so it is included as a lower confidence BL Lac candidate, though \citeauthor{hall02}\ have suggested it may be a BALQSO of extreme type.
{\it Top right:} SDSS~J220445.26$+$003141.9 ($z= 1.353$), published in \citeauthor{hall02}  This object again formally passes our selection criteria, and so it is included in our final catalog as a lower confidence BL Lac candidate.
{\it Bottom left:} SDSS~J130941.36$+$112540.1 ($z= 1.362$).  This new object appears similar, but was ultimately rejected as a BL Lac based on emission line strength.
{\it Bottom right:} SDSS~J145045.56$+$461504.2 ($z= 1.877$).  This new object was also rejected based on emission line strength.
}
\label{fig:oddbals}
\end{figure}

Interestingly, both the \citeauthor{hall02}\ objects, SDSS~J0105-0033 and SDSS~J2204+0031, were recovered in our database query, and they formally satisfy all our BL~Lac selection criteria.  Our database query also revealed two additional radio sources whose SDSS spectra appear very similar and may be related: SDSS J130941.36+112540.1 ($z$$=$1.362) and SDSS J145045.56+461504.2 ($z$$=$1.877). Both are again FIRST sources, and again neither new object matches to a RASS source.  The SDSS spectra of all four objects are shown in Figure \ref{fig:oddbals}, and Table \ref{tab:oddbal} displays some basic properties.  Redshifts for SDSS~J0105-0033 and SDSS~J2204+0031 are taken from \citeauthor{hall02}, while the redshifts for SDSS~J1309+1125 and SDSS~J1450+4615 are estimates based on \ion{Mg}{2} absorption. 

Neither of the two new cases, SDSS~J1309+1125 and SDSS~J1450+4615, ultimately survive as BL Lac candidates during manual inspection: they both appear to show broad emission features near \ion{Al}{3}/\ion{C}{3}.  We hence exclude the two new cases from our final BL~Lac catalog, although we retain  the original two \citeauthor{hall02}\ objects  as lower confidence BL Lac candidates.  Whatever their nature,
this set of radio objects are almost certainly AGN and very rare.

%%%%%%%%%%%%%%%%%%%%%
%%%%Mysterious Hall et al BALs%%%%

%%Odd BAL TABLE%%
\begin{deluxetable*}{cccccccc}
\tablewidth{0pt}
\tablecaption{Active Galactic Nuclei with Unusual Spectra\label{tab:oddbal}}
\tablehead{
	\colhead{SDSS Name} %1
	& \colhead{redshift} %2
	& \colhead{$u$} %3
	& \colhead{$g$} %4
	& \colhead{$r$} %5
	& \colhead{$i$} %6
	& \colhead{$z$} %7
	& \colhead{Radio Flux} %8
\\
	\colhead{(J2000)}       %1 
	&   %2
	& \colhead{(mag)} %3
	& \colhead{(mag)}  %4
	& \colhead{(mag)} %5     
	& \colhead{(mag)} %6       
	& \colhead{(mag)} %7     
	& \colhead{(mJy)} %8
}                
\startdata
010540.75$-$003313.9\tablenotemark{a} & 1.179 & 20.16 & 19.04 & 17.98 & 17.73 & 17.49 &  4.84 \\
130941.36$+$112540.1 & 1.362 & 21.23 & 19.75 & 18.57 & 17.89 & 17.82 &  0.83 \\
145045.56$+$461504.2 & 1.877 & 21.00 & 20.31 & 19.54 & 18.80 & 18.30 &  1.51 \\
220445.26$+$003141.9\tablenotemark{a} & 1.353 & 19.78 & 18.65 & 17.35 & 16.76 & 16.65 &  3.03 \\
\enddata
\tablenotetext{a}{Discovered by \citet{hall02}}
\end{deluxetable*}	
%%%%%%%%%%%%%%%%%%
%%%%%%%%%%%%%%%%%

%%%%%%%%%%%%%%%%%%%%%%%%%%%%%
%%%%%%%%%%%%%%%%%%%%%%%%%%%%%
\section{Final Sample}
\label{sec:catalog}
\begin{figure}
\centering
\includegraphics[scale=0.5]{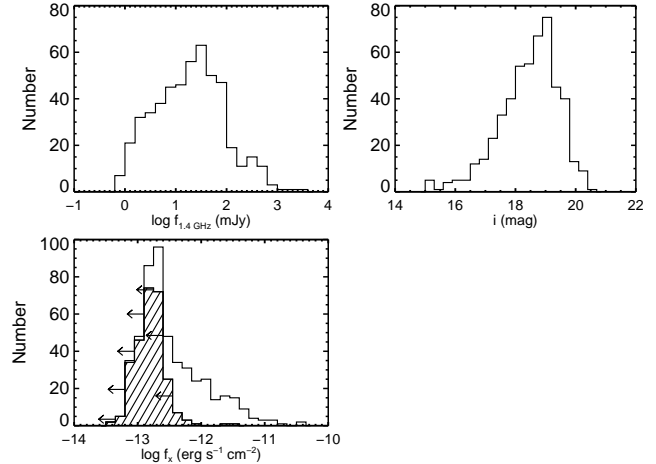}
\caption{Flux distributions for all \nsamp\ BL Lac candidates in our sample.  
{\it Top left:} Log radio flux densities (in mJy) from FIRST at 1.4 GHz.  
{\it Top right:} Optical magnitudes (sdss $i$).
{\it Bottom left:} Estimated log absorbed X-ray fluxes from 0.1-2.4 keV in erg~s$^{-1}$~cm$^{-2}$\ (see \S \ref{sec:fluxcalc}).  {\it Open histogram:} All \nsamp\ candidates.  {\it Shaded histogram:} Limits (denoted with arrows) for the 271 objects lacking RASS detections.
  }
\label{fig:fluxhist}
\end{figure}

Table \ref{tab:stat} summarizes the number of objects in our final sample, the number with X-ray detections in RASS, and some highlighted redshift information.  Thus far, from this and previous programs, the SDSS spectroscopic survey has already provided several hundred new BL Lac candidate identifications;  about 60\% of our sample were previously identified as BL Lacs by the SDSS (primarily in \citealt{collinge05} or \citealt{anderson07}), and $\sim$20\%\ of our sample were originally discovered by non-SDSS observations.  Only 33 of our BL Lac candidates appear as quasars in the DR5 Quasar Catalog \citep{schneider07}.  Although the DR5 Quasar Catalog does not include BL Lac objects, some overlap is expected between our BL Lac sample and the quasar catalog, as our BL Lac definition can include some very weak emission line objects.  Distributions of observed radio flux densities at 1.4 GHz (in mJy), observed optical magnitudes (sdss $i$), and estimated absorbed X-ray fluxes from 0.1-2.4 keV (in erg~s$^{-1}$~cm$^{-2}$, see \S \ref{sec:fluxcalc}) for all \nsamp\ BL Lac candidates are shown in Figure \ref{fig:fluxhist}.  

%%%%%%%%%%%%%%%%%%%%%
%%%%Flux Histograms%%%%

Data are given in Tables \ref{tab:empdata} and \ref{tab:deriveddata} for our \nsamp\  BL Lac candidates (\nbl\ higher confidence and \nmaybe\ lower confidence candidates.)  Table \ref{tab:empdata} presents primarily empirical information.  Column (1) gives the SDSS source name ({\it SDSS JHHMMSS.SS$\pm$DDMMSS.S}), sorted by right ascension.  Column (2) gives our classification as a higher confidence BL Lac (`BL') or as a lower confidence BL Lac (`BL?') candidate.  Column (3) notes whether each source is identified as a BL Lac in NED\footnote{The NASA/IPAC Extragalactic Database (NED) is operated by the Jet Propulsion Laboratory, California Institute of Technology, under contract with the National Aeronautics and Space Administration.}. `S' means the source is listed as a BL Lac discovered by SDSS, `O' means the source is listed as a BL Lac discovered by a survey other than SDSS, and `N' means it is not listed as a BL Lac in NED.  Columns (4) and (5) give the SDSS coordinates RA and Dec (J2000).  Columns (6)-(10) present $u$,$g$,$r$,$i$,$z$~psf magnitudes.  Column (11) gives the estimated redshift of each object (unknown redshifts are left blank), and column (12) presents our redshift flag:  `R' means we believe the redshift to be reliable, `T' is a tentative redshift, `L' means the redshift is derived from foreground absorbing gas and should be interpreted as a lower limit, and `U' designates an unknown redshift.  Column (13) gives each object's measured H/K break depression C as derived by us (not the SDSS pipeline measure.)   Column (14) gives the FIRST integrated radio flux listed in the SDSS DR5 database in mJy \citep{becker95,white97}.  Finally, additional comments may appear in column (15).  We explain in column (15) why each lower confidence candidate is classified as `BL?': `lowSN' signifies the optical spectrum has measured signal-to-noise $100< S/N\le 125$; `contam?' means the spectrum might be contaminated by a nearby source; and `ew' marks the objects with unclear continua, and emission lines can have measured $REW$\ larger or smaller than 5~\AA\ depending on where one chooses to measure the continuum near each line.  Not all SDSS discovered BL Lac candidates from \citet{collinge05} and \citet{anderson07} are identified as BL Lacs in NED.  We thus mark such objects in column (3) as not being listed in NED as a BL Lac, but we identify all objects that appear in  \citeauthor{collinge05}\ and/or \citeauthor{anderson07}\ as `C05' and `A07', respectively, in column (15).

Table \ref{tab:deriveddata} emphasizes derived multiwavelength information for each object (see \S \ref{sec:fluxcalc}).  We repeat SDSS source names and redshifts in columns (1)  and (2) for easy reference to Table \ref{tab:empdata}.  The 1.4 GHz FIRST integrated radio flux density (in mJy) is repeated in column (3), with the corresponding 5 GHz rest-frame monochromatic luminosity in column (4).  Column (5) gives the extinction corrected optical magnitude in the SDSS filter whose center is closest to rest-frame 5000 \AA, taken from the $r$\ filter for objects with unknown redshift.  Column (6) gives the estimated monochromatic optical luminosity at rest-frame 5000 \AA.   Column (7) is an X-ray match flag, where `Y' means there is a match to a RASS source, and `N' means there is no match.  Column (8) gives the X-ray count rate (in counts s$^{-1}$) from 0.1-2.4 keV \citep{voges99}.   The estimated monochromatic unabsorbed X-ray luminosity at rest-frame 1 keV is given in column (9).  All monochromatic luminosities have units erg s$^{-1}$~Hz$^{-1}$.  The X-ray count rates and luminosities presented in columns (8) and (9) are upper limits if the X-ray match flag in column (7) is set to {\it N}.    Finally, columns (10), (11), and (12) give the logarithmic broad-band spectral indices $\alpha_{ro}$, $\alpha_{ox}$, and $\alpha_{rx}$~respectively.  Lower limits to $\alpha_{ox}$~and $\alpha_{rx}$~are given if the X-ray match flag is set to `N'.  

As discussed in \S\ref{sec:queryeff}, 171 of the \citet{collinge05} optically selected probable BL Lac candidates that match to a FIRST radio source are recovered by our DR5 database query and additionally pass our signal-to-noise cut.  Two of these objects however are not included in our final  sample.  One object is rejected because it shows broad H$\alpha$\ in an H$\alpha$/[\ion{N}{2}] complex with  blended $REW$\ slightly larger than 5~\AA.  The second object is rejected because it shows \ion{Si}{4}+\ion{O}{4}\ with $REW$\ slightly larger than 5~\AA.  

The \citet{anderson07} X-ray selected sample contains 244 objects that match to a FIRST source in the DR5 spectroscopic database.  (After a cross-correlation to NVSS, only 7 objects in the \citeauthor{anderson07}\ sample lack radio detections.)  Our database query recovers 241 of these objects, of which 223 additionally pass our $S/N$ cut.  Only 2 of these 223 objects do not appear in our final radio-selected sample.  Both these objects were rejected here based on [\ion{O}{2}] emission with $REW$\ slightly larger than 5 \AA.  Thus, our final sample reassuringly contains essentially all FIRST sources previously identified as BL Lac candidates from SDSS spectroscopy.

%%%%%%%%%%%%%%%%%%%%%%%%%%%
%%%%%%Basic Sample Statistics Table%%%%%%%
\begin{deluxetable}{lccc}
%\tabletypesize{\small}
%\rotate
\tablewidth{0pt}
\tablecaption{ Basic Sample Statistics \label{tab:stat}}
\tablehead{
	\colhead{\nodata} %1
	& \colhead{Number of BL} %2
	& \colhead{Number of BL?} %3
	& \colhead{Total} %4
}
\startdata
Number of Objects & 426 & 75  & 501 \\
Match to RASS Source & 217 & 13 & 230 \\
Reliable Redshifts & 166 & 29 & 195 \\
Tentative Redshifts & 58 & 17 & 75 \\
Lower Limit Redshifts & 32 & 5  & 37 \\
Unknown Redshifts & 170 & 24 & 194  \\
\enddata
\end{deluxetable}

%%Emprical Info Table%%%
\begin{deluxetable*}{lccrrccccccccrl}
\tabletypesize{\scriptsize}
%\rotate
\tablewidth{0pt}
\tablecaption{Observed Parameters of BL Lac Candidates\label{tab:empdata}}
\tablehead{
	   \colhead{SDSS Name} %1
	&	%2	                
	& \colhead{in}%3
	& \colhead{RA} %4
	& \colhead{Dec} %5	     			     
	& %6		       
	& %7	 	         
	& %8	         
         & %9           
         & %10		  
         & %11	 	             
         & \colhead{Red.}%12		                                
         & %13	       
         &\colhead{$f_{1.4\ GHz}$} %14
         &  %15
\\
	   \colhead{(J2000)}  %1
	& \colhead{Class.}  %2
	& \colhead{NED? } %3 
	& \colhead{(J2000)}   %4
	& \colhead{(J2000)} %5
	& \colhead{$u$} %6 
	& \colhead{$g$} %7
	& \colhead{$r$} %8
	& \colhead{$i$} %9
	& \colhead{$z$} %10
	& \colhead{Redshift} %11
	& \colhead{Flag} %12
	& \colhead{$C$} %13
	& \colhead{(mJy)} %14
	& \colhead {Comm.}%15
\\
	 \colhead{(1)}      
	 & \colhead{(2)}      
	 & \colhead{(3)} 
	 & \colhead{(4)}  
	 & \colhead{(5)}  
	 & \colhead{(6)}  
	 & \colhead{(7)} 
	 & \colhead{(8)} 
	 & \colhead{(9)}  
	 & \colhead{(10)}  
	 & \colhead{(11)} 
	 & \colhead{(12)}    
	 & \colhead{(13)} 
	 & \colhead{(14)}
	 & \colhead{(15)}
}
\startdata
000157.23$-$103117.3 & BL & N &    0.48850 &  -10.52148 &  20.66 &  19.74 &  18.77 &  18.26 &  17.86 & 0.252 & R & 0.367 &   32.97 & \nodata   \\
000257.17$-$002447.3 & BL? & N &    0.73824 &   -0.41315 &  21.10 &  20.63 &  19.89 &  19.22 &  18.85 & 0.523 & R & 0.289 &  159.13 & lowSN   \\
002142.25$-$090044.4 & BL & N &    5.42608 &   -9.01234 &  20.15 &  19.79 &  19.32 &  18.82 &  18.52 & 0.648 & T & 0.217 &   41.96 &  C05  \\
002200.95$+$000657.9 & BL & O &    5.50396 &    0.11610 &  20.48 &  20.04 &  19.28 &  18.87 &  18.55 & 0.306 & R & 0.357 &    1.73 & C05,A07   \\
003808.50$+$001336.5 & BL & N &    9.53543 &    0.22683 &  20.15 &  19.70 &  19.30 &  18.96 &  18.63 & \nodata & U & \nodata &   89.71 &  C05  \\
\enddata
\tablecomments{The complete version of this table is in the electronic edition of the journal. The printed version contains only a sample.}
\end{deluxetable*}

%%Derived Info Table%%%
\begin{deluxetable*}{lcrccccccccc}
\tabletypesize{\scriptsize}
%\rotate
\tablewidth{0pt}
\tablecaption{Derived Parameters of BL Lac Candidates\label{tab:deriveddata}}
\tablehead{
	    \colhead{SDSS Name} %1
	& %2	     		               
	& \colhead{$f_{1.4\ GHz}$} %3
   	& %4			          
	&  %5                                   
	&  %6                                  	 
	& \colhead{X-ray} %7 
	& \colhead{X-ray Counts}  %8         
  	& %9				  
	& %10				     
	 & %11                                           
	 &  %12   
\\			   
	    \colhead{(J2000)}  %1 
	& \colhead{Redshift} %2
	& \colhead{(mJy)}        %3
	&  \colhead{$\log L_{5\ GHz}$} %4
	&  \colhead{$m_0$}  %5
	&  \colhead{$\log L_{5000~{\rm \AA}}$}   %6
	&  \colhead{Match Flag}   %7   
	&  \colhead{per Sec} %8
	&  \colhead{$\log L_{1\ keV}$}  %9
	&  \colhead{$\alpha_{ro}$}  %10
	&  \colhead{$\alpha_{ox}$}  %11 
	&  \colhead{$\alpha_{rx}$} %12
\\
	    \colhead{(1)}    
	 & \colhead{(2)}      
	 &   \colhead{(3)} 
	 &   \colhead{(4)}  
	 &   \colhead{(5)}  
	 &   \colhead{(6)}  
	 &   \colhead{(7)} 
	 &   \colhead{(8)} 
	 &   \colhead{(9)}  
	 &   \colhead{(10)}  
	 &  \colhead{(11)} 
	 &  \colhead{(12)} 
}
\startdata
000157.23$-$103117.3 & 0.252 &   32.97 &  31.82 &  18.65 &  29.29 & N &   0.019 &  25.91 &   0.50 &   1.30 &   0.77 \\
000257.17$-$002447.3 & 0.523 &  159.13 &  33.15 &  19.15 &  29.76 & N &   0.015 &  26.63 &   0.67 &   1.20 &   0.85 \\
002142.25$-$090044.4 & 0.648 &   41.96 &  32.75 &  18.46 &  30.16 & N &   0.018 &  26.94 &   0.51 &   1.24 &   0.76 \\
002200.95$+$000657.9 & 0.306 &    1.73 &  30.71 &  19.21 &  29.27 & Y &   0.097 &  26.85 &   0.28 &   0.93 &   0.50 \\
003808.50$+$001336.5 & \nodata &   89.71 &  32.40 &  19.25 &  29.23 & N &   0.019 &  26.02 &   0.63 &   1.23 &   0.83 \\
\enddata
\tablecomments{The complete version of this table is in the electronic edition of the journal. The printed version contains only a sample.}
\end{deluxetable*}
%%%%%%%%%%%%%%%%%%%%%%%%
%%%%%%%%%%%%%%%%%%%%%%%%%%

%%%%%%%%%%%%%%%%%%%%%%%%%%%
%%%%%%%%%%%%%%%%%%%%%%%%%%%
\section{Discussion}
\label{sec:disc}
%%%%%%%%%%%%%%%%%%%%%%
The following analysis conservatively focuses on the \nbl\ higher confidence BL Lac candidates.  Results remain virtually unchanged when all \nsamp\ candidates are included.  

\begin{figure}
\centering
\includegraphics[scale=0.50]{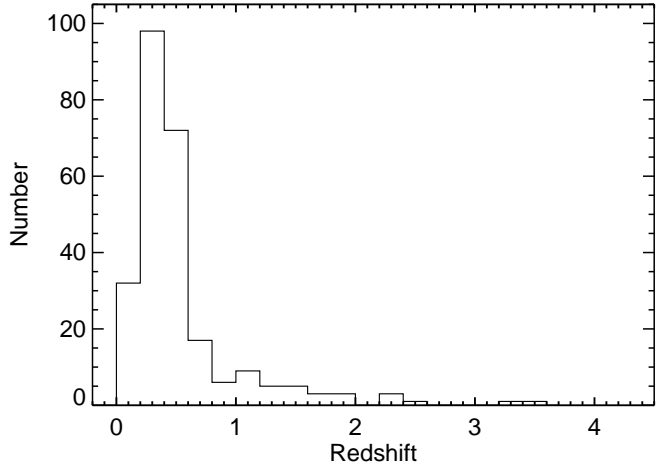}
\caption{Redshift distribution of the 256 higher confidence candidates with reliable, tentative or lower limit redshifts.  All redshifts with $z<1.1$\ are derived from host galaxy features.  The relative number of objects with measured redshifts decreases drastically for redshifts larger than $z\sim0.5$.  However, a sizable fraction (30\%) still have measured redshifts larger than 0.5, making this a useful sample for constraining higher redshift BL Lac properties.}
\label{fig:redshift}
\end{figure}

Figure~\ref{fig:redshift} shows a histogram for the 256 higher confidence BL Lac candidates with either reliable, tentative or lower limit redshifts.    A sizable fraction (30\%) have redshifts larger than 0.5 (18\% with $0.5 < z < 1.0$), making this a useful sample for testing BL Lac properties at $z>0.5$.   All objects with $z<1.1$\ have a measured \ion{Ca}{2}~H/K break, but note that we observe a sharp decrease at $z\sim0.5$\ in the distribution of measured redshifts.   It is possible that some objects with unknown redshifts are examples of highly beamed nearby BL Lacs; however, given the likelihood of detecting host galaxy spectral features from lower redshift objects, it is probable that a significant fraction of these objects are higher redshift between $0.5 \lesssim z \lesssim 2.2$.  (Spectra with $z\gtrsim 2.2$\ would start to show Ly$\alpha$\ forests.)  Also, highly beamed objects require a low probability geometry and may therefore be less likely to be recovered at lower redshifts where a smaller volume is probed.  

%%%%%%%%%%%%%%%%%%%%
%%%%%%%Redshift%%%
%%%%%%%%%%%%%%%%%%
%%%%%%%%%%%%%%%%%%%

\begin{figure}
\centering
\includegraphics[scale=0.50]{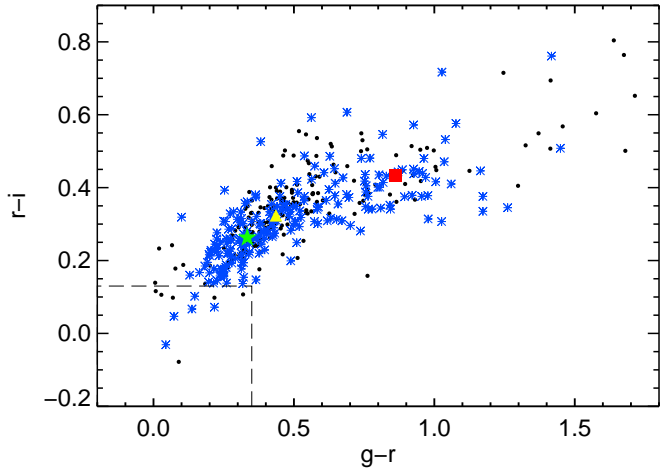}
\caption{Color-color diagram of our higher confidence BL Lacs.
{\it Black circles} show higher confidence BL Lac candidates that do not match to a RASS X-ray source. 
{\it Blue asterisks} show higher confidence BL Lac candidates within 1$'$\ of a RASS source.   We recover X-ray detections at all ranges of optical colors. 
The {\it green star} depicts the median colors of objects not showing a \ion{Ca}{2}\ H/K break in their spectra.  
The {\it yellow triangle} marks the median colors of objects with small measured H/K breaks $C\le0.25$.  
The {\it red square} indicates the median colors of objects showing larger H/K breaks $0.25<C\le0.4$.  As expected, objects with more prominent host galaxy features show redder optical colors on average.  
The {\it dashed lines} mark the ``{\it gri} box'' identified by \citet{collinge05} where most stellar contaminants would be expected.  Eight of the 11 objects recovered in this box have measured redshifts, supporting our conclusion of little stellar contamination to our sample.   
Three objects with $g-r>1.8$\ and/or $r-i>0.9$\ due to poor flux measurements are omitted from this figure.}
\label{fig:optcol}
\end{figure}

A color-color diagram for our \nbl\ higher confidence BL Lac candidates is shown in Figure~\ref{fig:optcol}.    Objects with RASS X-ray detections are marked as blue asterisks, and those lacking X-ray detections are shown as black circles.  Note that X-ray detected and non-detected BL Lac candidates cover the same range of optical colors.  That is, the types of BL Lacs that can show X-ray emission are not limited to one specific subclass.   The dashed lines mark the region ($g$$-$$r < 0.35$, $r$$-$$i < 0.13$) where \citet{collinge05} found most stellar contaminants in their optically selected BL Lac sample.    We find only 11 objects in this ``{\it gri} box''; none show significant proper motion and 8 have tentative or reliable extragalactic redshifts.  This affords added confidence that any potential stellar contamination to our radio sample is extremely small, and it supports the notion that requiring multifrequency information improves BL Lac selection efficiency.  However, we note that multifrequency approaches are not as capable of recovering potentially rare (if extant) and extremely interesting BL Lac sub-populations such as, for example, radio-quiet BL Lacs.

%%%%%%%%%%%%%%%%%%%%%%
%%%%Optical Color-color plots%%%%%%%
%%%%%%%%%%%%%%%%%
%%%%%%%%%%%%%%%%%%

\begin{figure}
\centering
\includegraphics[scale=0.50]{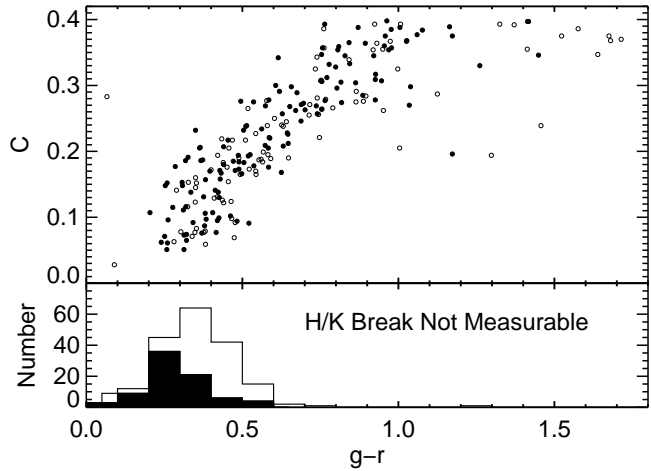}
\caption{\ion{Ca}{2} H/K depression strength $C$\ vs.\ $g-r$ for our higher confidence candidates.  
{\it Top panel:}  objects (234) with a measured H/K break.
The {\it filled circles} depict objects that match to a RASS X-ray source,
while the {\it open circles} show those without an X-ray match.  
{\it Bottom panel:} objects without a measured H/K break either because 4000~\AA\ rest-fame is redshifted out of the SDSS spectra or because the objects have unknown redshifts. 
{\it Open histogram:} all 192 higher confidence candidates lacking a measured H/K break.
The {\it shaded histogram} shows the 79 objects with an X-ray detection in RASS that also lack H/K break measurements.
Objects with more prominent host galaxy contamination (i.e., larger $C$) tend to be redder, as expected in the standard beaming paradigm for the BL Lac phenomenon.  Also note that objects with X-ray detections are recovered at all ranges of $C$\ and $g-r$.  Two objects with $g-r>1.8$\ due to poor flux measurements are omitted from this plot.}
\label{fig:cVgr}
\end{figure}

Spectra showing larger \ion{Ca}{2} H/K depressions tend to have redder optical colors on average.  The red square in Figure~\ref{fig:optcol} marks the median colors of objects whose spectra show larger \ion{Ca}{2} H/K depressions ($0.25<C\le0.4$), the yellow triangle marks the median colors of spectra showing smaller depressions ($0<C\le0.25$), and the green star marks the median colors of spectra that do not show H/K breaks (either because the break is completely washed out by the AGN flux or because the H/K break is redshifted out of the optical spectrum.)    This trend can be seen more directly by examining H/K break strength vs.\ color for our higher confidence candidates (Figure \ref{fig:cVgr}).  Our optical spectra are the combination of two distinct radiation sources: the host galaxy (typically an elliptical) and a typically blue power law continuum from the AGN.  In the standard BL Lac beaming paradigm, objects with their jets pointed less directly at the observer show less non-thermal continua (due to less Doppler boosting), and therefore they exhibit more prominent host galaxy features and redder optical colors.  The open histogram at the bottom of Figure \ref{fig:cVgr} shows all 192 higher confidence candidates for which we have no measured H/K break (either because the H/K break is redshifted out of the optical spectrum, or because the object has an unknown redshift so it is impossible to locate 4000~\AA\ rest-frame.)  The objects without a measured H/K break tend to be bluer on average than those with a larger break (e.g., $C \gtrsim 0.2$).  Also note in Figure \ref{fig:cVgr} that X-ray detected objects are shown as filled circles in the top panel and as a filled histogram in the bottom panel; we detect X-rays from objects at all ranges of $g$$-$$r$\ and H/K break strength.  

%%%%%%%%%%%%%%%%%%
%%%%H/K vs. color$$$$
%%%%%%%%%%%%%%%%%
%%%%%%%%%%%%%%%%%

\begin{figure}
\centering
\includegraphics[scale=0.50]{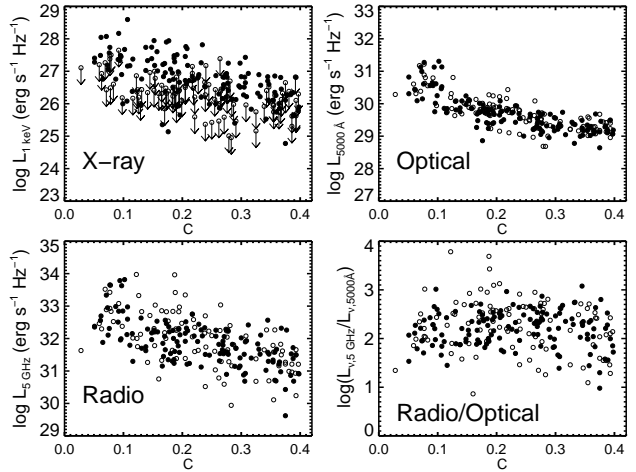}
\caption{ Multiwavelength luminosities vs.\ \ion{Ca}{2} H/K break strength for our higher confidence candidates.  
{\it Filled circles}: objects that match to a RASS X-ray source.
{\it Open circles}: objects without an X-ray match (with arrows designating X-ray limits when appropriate.)
{\it Top left:} logarithm of specific X-ray luminosity at 1~keV vs.\ \ion{Ca}{2} H/K break strength $C$.
{\it Top right:} logarithm of specific optical luminosity at 5000~\AA\ vs.\ $C$.
{\it Bottom left:} logarithm of specific radio luminosity at 5~GHz vs.\ $C$.  BL Lacs tend to appear more luminous at all observed frequencies when viewed more directly along their jet axes (i.e., for smaller values of $C$).
{\it Bottom right:} logarithm of radio to optical specific luminosity ratio vs.\ $C$.  The large range of radio to optical luminosity ratios does not appear to be an orientation effect and rather probably reflects AGN jet physics (e.g., variability and/or differing SEDs among BL Lac objects.) 
Note that we obtain X-ray detections at all values of H/K break strength and radio and optical luminosities. 
} 
\label{fig:hkplts}
\end{figure}

The standard beaming paradigm is further supported through an anti-correlation between H/K break strength and the log specific luminosity, shown in Figure~\ref{fig:hkplts} for the X-ray (top left), optical (top right), and radio (bottom left).  We find linear Pearson correlation coefficients $r= -0.60$\ and $r=-0.78$\ for the logarithm of radio and optical luminosities respectively; both anti-correlations are significant at the $>$99.999\% level.   That is, BL Lacs appear to be more luminous as we view them more directly along their jet axes.   Even though there is an observed trend, the large scatter in Figure~\ref{fig:hkplts} would seem to prohibit the use of H/K break as a sole proxy for predicting accurate luminosities.  Also, while our sample generally supports  the beaming scenario, we note that it is still possible that a minority of sources might have alternative explanations for exhibiting BL Lac type characteristics.   Detailed investigations of such a possible (albeit likely rare) sub-population are beyond the scope of this paper. 

We do not show the luminosity distributions in Figure~\ref{fig:hkplts} for objects without a measured H/K break, as only 22 such objects have measured redshifts.  The average X-ray, optical and radio luminosities of 10$^{28.0}$, 10$^{31.4}$, and 10$^{32.9}$ (erg~s$^{-1}$~Hz$^{-1}$) for these 22 objects however do lie toward toward the bright end of their respective luminosity distributions, as expected.  We note though that all 22 objects are high redshift ($z>1.222$) and therefore may be observationally biased toward higher luminosities.   

The scatter in optical luminosity vs.~H/K break  is smaller than in the radio or X-ray, as is expected for two optical measures.  BL Lacs can have diverse SEDs, as the peak of the synchrotron component can fall within a range spanning close to 10 orders of magnitude in frequency \citep[e.g., see][]{nieppola06}.  For example, two BL Lac jets with identical optical brightnesses can have very different radio (or X-ray) luminosities depending on the specific shape of each object's SED.  This plausibly introduces excess scatter into the radio and X-ray panels in Figure~\ref{fig:hkplts}, especially since we use the same spectral indices for all objects when estimating luminosities.  Also, given the non-simultaneity of our multifrequency data, variability is likely responsible for some of the observed scatter.  

%%%%%%%%%%%%%%%%
%%%%Luminosity vs. H/K break%%%%
%%%%%%%%%%%%%%%%%
%%%%%%%%%%%%%%%

The bottom right panel of Figure~\ref{fig:hkplts} shows the log of the ratio of radio to optical specific luminosity vs.\ measured H/K break.   Here, we observe no obvious linear trend (a 28\% two-sided  probability of randomly finding a stronger correlation), but we do see a relatively large scatter.  The large scatter likely confirms either variability and/or that BL Lacs do not have a universal SED.  We omit objects lacking H/K break measurements for consistency with the other panels.  The lack of an obvious observed trend with H/K break strength suggests that the large observed range of radio to optical luminosity ratios is not a strong orientation effect and perhaps reflects AGN jet physics.

The bottom right panel of Figure \ref{fig:hkplts} also illustrates one reason that multifrequency surveys should be well matched in depth.  A radio survey that is shallower than an optical survey, for example, will be biased toward recovering objects with larger radio to optical luminosity ratios (i.e., extreme LBLs).   One needs {\it both} a deep radio and a deep optical survey to recover objects toward the bottom of the bottom right panel of  Figure~\ref{fig:hkplts}\ and to obtain a sample with luminosity ratio distributions more representative of the underlying population.  

%%%%%%%%%%%%%%%%%%%
%%%%%%%%%%%%%%%%%%%%

\begin{figure}
\centering
\includegraphics[scale=.5]{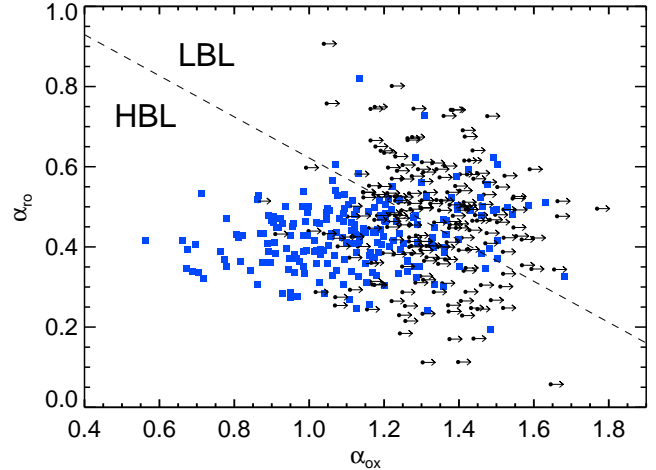}
\caption{Broad-band spectral indices $\alpha_{ro}$\ vs.\ $\alpha_{ox}$.  Radio brightness increases toward the top, and X-ray brightness increases toward the left.
The {\it blue squares} show objects that match to a RASS X-ray source
while the {\it filled circles} depict objects without an X-ray match (with arrows designating limits).
We again obtain X-ray detections at essentially all ranges of parameters (with the radio weak $\alpha_{ro}$\ tail being the notable exception).  We observe a continuous distribution of broad-band spectral indices with plenty of IBL and LBL-type objects.  However, we do see a potential bias toward recovering HBLs ($\alpha_{rx} \le 0.75$) from SDSS spectroscopy, a perhaps unexpected result for a radio-selected sample.  This is maybe an artifact from the SDSS spectroscopic target algorithms.  
}

\label{fig:alpha}
\end{figure}

Figures \ref{fig:optcol}-\ref{fig:hkplts} demonstrate that BL Lacs display continuous distributions in their properties, supporting the notion that the observed dichotomy between radio and X-ray selected BL Lacs in the 1 Jy and EMSS samples was indeed plausibly a selection effect induced by small sample size and bright survey flux limits.  This is further supported when examining the $\alpha_{ro}-\alpha_{ox}$~plane in Figure \ref{fig:alpha} (which shows our \nbl\ higher confidence BL Lac candidates.)  Blue squares represent objects with X-ray detections; arrows designate limits on $\alpha_{ox}$~for objects lacking a RASS detection.  Here, large (small) values of $\alpha_{ro}$~($\alpha_{ox}$) correspond to being radio (X-ray) bright.  As expected, we observe a continuous transition between LBL and HBL objects\footnote{HBLs are commonly defined as $\alpha_{rx}\le0.75$ corresponding to $F_x/F_r\gtrsim10^{-11}$~with the X-ray flux in erg s$^{-1}$~cm$^{-2}$~and the radio flux in Jy \citep{padovani95_apj}.  The dividing line is also sometimes quoted as the similar $F_x/F_r\ge10^{-5.5}$~where both fluxes are in Jy \citep{wurtz94,perlman96}.}, with plenty of IBL-type objects bridging the transition.   Although this preliminary analysis seems to support the notion of a single BL Lac population, there are still some differences between historically radio selected and X-ray selected BL Lacs that require further attention.  For example, XBLs appear to have more starlight contamination from their host galaxies, lower polarization levels, and constant polarization position angles (see \S\ref{sec:intro}).  Also, XBLs and RBLs might have very different cosmological evolution. 

We see an obvious and expected bias toward RASS detected sources preferentially filling the HBL parameter space in Figure \ref{fig:alpha}:  83\% of X-ray emitters appear to be HBLs, likely due to the relatively high flux limit of the RASS survey.   Perhaps counter intuitively however, 66\% of our entire higher confidence radio sample (and 47\% of objects lacking RASS detections) still have broad-band colors indicative of HBLs.  This may be an artifact of the SDSS spectroscopic target selection algorithms, which are biased toward selecting objects near RASS  sources for spectroscopy (see \S\ref{sec:sampsel} and Figure \ref{fig:sed}).  Keep in mind however that the above percentages are tentative for several reasons.  About half our data points only have limits on their X-ray luminosites; the non-simultaneity of our multiwavelength data and our monochromatic luminosity estimates also introduce excess scatter.   

We also note X-ray detections from some LBLs in Figure \ref{fig:alpha}.  That is, as is similarly observed in Figures~\ref{fig:optcol}-\ref{fig:hkplts}, X-ray emission is detected from objects at nearly all ranges of observed parameters (objects with $\alpha_{ro} \lesssim 0.2$~being the notable exception, see below.)   While it is certainly plausible that some fraction of radio bright objects lacking X-ray detections are real examples of extreme LBLs, the recovery of X-ray sources at all parameters suggests that we do not have evidence for a significant population of X-ray quiet BL Lacs.  Deeper X-ray observations are necessary for testing this assertion, and selection effects and survey biases need to be examined thoroughly. 

There is a small population of 7 higher confidence candidates with extremely small $\alpha_{ro}<0.2$, of which only a single (low-redshift) object shows X-ray emission, implying that some BL Lacs might have atypical broad-band colors.  However, as described in \S \ref{sec:qso}, five of these objects have redshifts $z>1$, potentially making their classification as BL Lac candidates suspicious.  The other 26 objects in our higher confidence BL Lac candidate sample with $z>1$, however, do not show atypical broad band colors.  We include all $z>1$\ objects in our sample without prejudice, as all formally pass our BL Lac criteria; follow-on polarization, variability, and multiwavelength studies of these higher redshift BL Lac candidates (or weak lined quasars) would be useful.

%%%%%%%%%%%%%%%%%%
%%%%%alpha--alpha%%%%%%%%
%%%%%%%%%%%%%%%%%%%%%%%
%%%%%%%%%%%%%%%%%%%

%%%%%%%%%%%%%%%%%%%%%%%%%%%%%%
Looking forward to future applications, the very large size of our homogeneous SDSS/FIRST selected sample may (when coupled with a detailed understanding of selection issues) permit a number of subsequent BL Lac studies ranging from the potentially unusual evolution of the luminosity function of BL Lacs \citep[e.g.,][]{morris91,wolter94, bade98, rector00, rector01, giommi01, beckmann03, padovani07}, to tests of the unification paradigm.
As just one specific future application, consider the standard model that interprets BL Lacs as FR I radio galaxies with their jets pointed toward the observer.  Ground and space-based observations that resolve BL Lac host galaxies tend to support this unification scheme at least for the majority of objects.  An HST survey (primarily with WFPC2 SNAP observations) found strong direct evidence at lower redshifts supporting this scheme \citep{scarpa00_hsti, urry00_hstii, falomo00_hstiii, scarpa00_hstiv}.  However, it is not yet clear whether it holds for every single BL Lac.  Suggested modifications range from some fraction of BL Lacs being hosted by FR II radio galaxies \citep[e.g.,][]{owen96,wurtz97,rector01,hardcastle03} to some fraction being microlensed \citep[e.g.,][]{ostriker90,stocke97,rector01}; any requisite modifications are likely more significant at higher redshifts. Moreover, some caution may be appropriate when extrapolating the Urry et al.\ HST survey results to higher redshifts, as only 6 host galaxies with $z$$>$0.5 were actually resolved due to limited WFPC2 SNAP exposure times. Even if one conservatively disregards our \nmaybe\ lower confidence BL Lac candidates and all BL Lac candidates with $z>1$, our sample still contains a sizable number (46) of higher confidence BL Lac candidates with confirmed $0.5<z<1$\ suitable for host galaxy investigations.  If observed with current and upcoming {\it HST} instruments, our sample could potentially provide an important extension to the $0.5<z<1$\ regime of the well-established low redshift unification results.

%%%%%%%%%%%%%%%%%%%%%%%%%
%%%%%%%%%%%%%%%%%%%%
\section{Summary}
\label{sec:summary}
We have assembled a \nsamp\ object sample of BL Lac candidates selected jointly from SDSS DR5 optical spectroscopy and FIRST radio imaging: there are \nbl\ higher confidence candidates and \nmaybe\ lower confidence candidates.  All \nsamp\ objects emerge with homogeneous multi-wavelength data coverage from FIRST, SDSS and RASS.  After imposing a signal to noise cut on the SDSS spectra, we require all emission features to have $REW<5$\ \AA\  and, if present,  a \ion{Ca}{2} H/K depression $C\le0.40$.  Special attention is given to remove radio galaxies, radio stars, broad absorption line quasars, and other weak-featured objects that might otherwise formally pass our selection criteria.  Of particular interest, in addition to recovering 2 unusual radio AGN reported previously in \citet{hall02}, we found 2 additional objects possibly of the same class.   

We recover essentially all previous SDSS selected BL Lac candidates that match to FIRST sources in \citet{collinge05} and \citet{anderson07}, increasing confidence in our selection technique.  Of our \nsamp\ BL Lac candidates,  approximately 80\%\ are SDSS discoveries (including objects first reported by \citealt{anderson03}, \citealt{collinge05}, or \citealt{anderson07}).  

Approximately 60\% of our \nsamp\ BL Lac candidates have either a reliable redshift (derived from at least 2 spectral features at the same redshift), a tentative redshift (derived from a single emission line assumed to be \ion{Mg}{2}, or from multiple weak spectral features at the same redshift), or a redshift lower limit derived from common intervening absorption line doublets.  Of our 256 higher confidence candidates with spectroscopic redshifts measured from SDSS spectroscopy, 30\% have $z>0.5$, and 18\% have $0.5<z<1$, making this an excellent sample for constraining higher redshift BL Lac properties.

Almost half of our higher confidence candidates match within $1'$ to a RASS X-ray source.  We find X-ray detections for BL Lacs at practically all ranges of parameter space (i.e., optical colors, multi-wavelength luminosities, H/K break strengths, etc.), suggesting that there is not a significant population of X-ray quiet BL Lacs.  Perhaps surprisingly, only about one-third of our higher confidence radio-selected candidates have broad-band spectral indices indicative of LBLs.  This may be an artifact of the SDSS spectroscopy targeting algorithms that could bias spectroscopic observations in favor of objects near X-ray sources.  However, we still recover a large number of IBL/LBL-type objects; follow-up of this sample will therefore probe a wide range of BL Lac physics.

Our sample supports the general concurrence that BL Lacs display a continuous distribution of properties, but we stress that the historical differences between XBLs and RBLs still need to be accounted for and require further attention.   A preliminary analysis seems consistent with the standard beaming model, in that more highly beamed (i.e., more luminous) objects show weaker host galaxy features.  We also see indirect evidence for variability and/or diverse SED shapes, and we note that multi-wavelength surveys should be well-matched in depth to recover unbiased BL Lac samples. 

Our SDSS/FIRST BL Lac sample constitutes one of the largest homogeneous samples to date, with the attractive attribute that all objects emerge with the same high-quality radio, optical, and X-ray data. A thorough investigation of the selection effects that shape this sample's composition may allow follow-on studies to place useful constraints on BL Lac properties ranging from evolution to unification models, especially extending to the high-redshift tail of the BL Lac population.

%%%%%%%%%%%%%%%%
\acknowledgments
We gratefully thank the referee for providing excellent suggestions for improving this manuscript.  R.M. Plotkin and S.F. Anderson acknowledge support
from NASA/ADP grant NNG05GC45G. Funding for the SDSS and SDSS-II 
has been provided by the Alfred P. 
Sloan Foundation, the Participating Institutions, the National Science 
Foundation, the U.S. Department of Energy, the National Aeronautics and 
Space Administration, the Japanese Monbukagakusho, the Max Planck Society, 
and the Higher Education Funding Council for England. The SDSS Web Site is 
http://www.sdss.org/.  The SDSS is managed by the Astrophysical Research Consortium for the 
Participating Institutions. The Participating Institutions are the 
American Museum of Natural History, Astrophysical Institute Potsdam, 
University of Basel, Cambridge University, Case Western Reserve 
University, University of Chicago, Drexel University, Fermilab, the 
Institute for Advanced Study, the Japan Participation Group, Johns Hopkins 
University, the Joint Institute for Nuclear Astrophysics, the Kavli 
Institute for Particle Astrophysics and Cosmology, the Korean Scientist 
Group, the Chinese Academy of Sciences (LAMOST), Los Alamos National 
Laboratory, the Max-Planck-Institute for Astronomy (MPIA), the 
Max-Planck-Institute for Astrophysics (MPA), New Mexico State University, 
Ohio State University, University of Pittsburgh, University of Portsmouth, 
Princeton University, the United States Naval Observatory, and the 
University of Washington.

%%%%%%%%%%%%%%%%%%%%%
%%%%%%%%%%%%%%%%%%%%%%
%\bibliography{/Users/rich/bllac_ref}

\end{document}